\documentclass[prb,twocolumn,superscriptaddress,showpacs]{revtex4}

\usepackage{graphicx} 
\usepackage{epsfig}   
\usepackage{amssymb,amsmath,amsfonts,hyperref}
\usepackage{wasysym}
\usepackage{latexsym}
\usepackage{color}

\newcommand{\la}{\langle}
\newcommand{\ra}{\rangle}

\newcommand{\avg}[1]{{\bf E}\left[ #1\right]}

\begin{document}

\title{Random Coulomb antiferromagnets: \\ from diluted spin liquids to Euclidean random matrices}
\author{J. Rehn}
\affiliation{Max-Planck-Institut f\"ur Physik komplexer Systeme, 01187 Dresden, Germany}
\author{Arnab Sen}
\affiliation{Department of Theoretical Physics, Indian Association for the Cultivation of Science, Kolkata 700032, India}
\affiliation{Max-Planck-Institut f\"ur Physik komplexer Systeme, 01187 Dresden, Germany}
\author{A. Andreanov}
\affiliation{Max-Planck-Institut f\"ur Physik komplexer Systeme, 01187 Dresden, Germany}
\author{Kedar Damle}
\affiliation{Department of Theoretical Physics, Tata Institute of Fundamental Research, Mumbai 400 005, India}
\author{R. Moessner}
\affiliation{Max-Planck-Institut f\"ur Physik komplexer Systeme, 01187 Dresden, Germany}
\author{A. Scardicchio}
\affiliation{Abdus Salam ICTP, Strada Costiera 11, 34151 Trieste, Italy}
\date{\today}

\begin{abstract}
We study a disordered classical Heisenberg magnet with uniformly antiferromagnetic
interactions which are frustrated on account of their long-range Coulomb form, {\em i.e.}
$J(r)\sim -A\ln r$ in $d=2$ and $J(r)\sim A/r$ in $d=3$. This arises naturally as the
$T\rightarrow 0$ limit of the emergent interactions between vacancy-induced degrees
of freedom in a class of diluted Coulomb spin liquids (including the classical
Heisenberg antiferromagnets on checkerboard, SCGO and pyrochlore lattices)
and presents a novel variant  of a disordered long-range spin Hamiltonian.
Using detailed analytical and numerical studies we establish that this model exhibits
a very broad paramagnetic regime that extends to very large values of $A$ in both
$d=2$ and $d=3$. In $d=2$, using the lattice-Green function based finite-size
regularization of the Coulomb potential (which corresponds naturally to the underlying
low-temperature limit of the emergent interactions between orphan-spins), we only find
evidence that freezing into a glassy state occurs in the limit of strong coupling, $A=\infty$,
while no such transition seems to exist at all in $d=3$.
We also demonstrate the presence and importance of screening for such a magnet.
We analyse the spectrum of the Euclidean random matrices describing a Gaussian version of
this problem, and identify a corresponding quantum mechanical scattering problem.
\end{abstract}

\pacs{
	xx
}
\maketitle

\section{{\bf INTRODUCTION}}
The appearance of novel magnetic phases\cite{Moessner_Ramirez,balents2010spin,binder1986spin} generally contains as one
ingredient the ability of the system to avoid conventional (semi-)classical ordering.
In this connection, the role of several factors has been extensively explored.
These include low dimensionality and the resulting enhancement in the effects
of quantum and entropic fluctuations, geometrical frustration, whereby the leading
antiferromagnetic interactions compete with each other on lattices such as the
kagome and pyrochlore lattice, and the presence of quenched disorder, which
disrupts any residual tendency to conventional long-range order. Each of these
has given rise to research efforts spanning decades of work.

Here, we study a model with a  new combination of some of these ingredients.
The focus of our study is a disordered classical Heisenberg magnet with antiferromagnetic
interactions which are frustrated on account of their long-range Coulomb form at
long-distances, {\em i.e.} $J(r)\sim -A\ln(r/{\mathcal L})$ in $d=2$ (where
${\mathcal L}$ is a length-scale of order the system-size) and $J(r)\sim A/r$
in $d=3$. This Coulomb form of the Heisenberg couplings arises naturally as the
$T\rightarrow 0$ limit of the emergent entropic exchange interactions~\cite{sen2012vacancy}
between vacancy-induced ``orphan-spin'' degrees of freedom
\cite{schiffer1997two,moessner1999magnetic,Henley2001,Henley2010coulomb} in
diluted Coulomb spin liquids, and presents a novel variant  of a disordered
long-range spin Hamiltonian with connections to Euclidean random matrices.
The coupling constant $A$ is determined in any given system by the microscopic
details of the underlying Coulomb-spin liquid, while the spin degrees of freedom in
the model we study are related to the physical orphan-spins of the underlying
diluted magnet. Our focus here is on studying the range of behaviours possible
in the $T\rightarrow 0$ limit by mapping out the phase diagram of our Coulomb
antiferomagnet as a function of $A$.
Frustration arises naturally in the model under consideration, as {\em any}
triplet of spins are mutually coupled antiferromagnetically but without the
randomness in sign of, say, the Sherrington-Kirkpatrick model~\cite{sherrington1975solvable}.
Also, unlike the latter case, the interactions are long-ranged but not independent
of distance.

Our motivations for studying it include having been led to this model in a previous
investigation~\cite{SenDamleMoessnerPRL} of diluted frustrated magnets exhibiting a
Coulomb spin liquid at low temperature. The model is in this sense natural,
appearing as the zero-temperature limit of a disordered frustrated magnet.
The corresponding experiments are on the material known as SCGO, which triggered
the interest in what we now call highly frustated magnetism in the late
80s~\cite{obradors1988magnetic}. Its behaviour at very low temperatures is
still not very well understood, e.g. the observed glassiness even at very low
impurity densities \cite{ramirez1990strong, laforge2013quasispin}, which
appears to involve only  the freezing of a  fraction of its degrees of freedom.
We will return to this point in Sec.~\ref{DiscFreezInFrust}.
While exhibiting a classical Coulomb  spin liquid regime, the disorder in
this system leads to the emergence of new, fractionalized, degrees of freedom,
the so-called Orphans~\cite{schiffer1997two,moessner1999magnetic}, which interact
via an  effective entropic long range interaction mediated by their host spin
liquid \cite{sen2012vacancy}.

We believe that as such, it can be of interest as a generic instance of the
interplay of strong interactions and disorder in magnetism. In particular,
it develops the strand of thought of how disorder in a topological system
characterised by an emergent gauge field can nucleate gauge-charged defects,
with the pristine bulk mediating an effective interaction between them.
Long-range Coulomb interactions like the one studied here are then as natural
as the algebraically decaying RKKY interactions in metallic spin glasses.

Our central results are the following. First we use the results of previous work\cite{sen2012vacancy}, 
to work out in detail the key features of this $T\rightarrow0$ limit, and demonstrate
that this limit is characterized by a single coupling constant $A$, which is, in principle,
determined by the geometry of the underlying spin-liquid. Second, our extensive Monte Carlo
simulations for $d=2$ reveal no sign of any freezing or ordering transition up to very
large coupling strengths. At the same time, within a self-consistent Gaussian approximation,
we find that there does appear such a transition at infinite coupling in $d=2$ but not
in $d=3$. This transition is very tenuous, in that it is replaced by a more conventional
ordering transition in a finite system depending on the choice of how to regularize this
long-range interaction in a finite lattice: the finite-size lattice
regularization that is most natural from the point of view of the $T\rightarrow0$ limit of the
underlying diluted magnet gives rise to freezing into a glassy state at $A^{-1} = 0$,
while other regularizations replace this glassy state by a conventional ordering pattern.
The Coulomb antiferromagnet therefore remains highly susceptible to
perturbations, just like many other frustrated magnets~\cite{Moessner_Ramirez}.

We also study the spectrum of the interaction matrix of this random Coulomb antiferromagnet, 
which provides an instance of an Euclidean random matrix~\cite{mezard1999spectra,goetschy2013euclidean}, in that its entries are obtained as
a distance function between randomly chosen location vectors. We find two qualitatively distinct
regimes. On one hand, at low energies in the low-density limit, eigenfunctions are localised, 
with the lowest energy states as pairs of neighbouring spins the probability distribution
of which we compute. Beyond this extreme low-density limit, more complex lattice animals
appear in this regime.
On the other hand, at high energies, the modes correspond to long-wavelength
charge density variations with superextensive energy. In between, we find no
clear signature of a well-defined mobility edge in this Coulomb system. 

Another interesting aspect of the uniformly antiferromagnetic interactions
is that they permit a  variant of screening to appear in this Coulomb magnet,
which has no correspondence with other long-range magnets such as the
Sherrington-Kirkpatrick model. Our analysis of this screening further leads
us to an identification of the correlations of the random Coulomb antiferromagnet
with the properties of the zero-energy eigenstate of a quantum particle
in a box with randomly placed scatterers.

Returning to experiments, we note that the uniform magnetic susceptibility
of SCGO will of course be dominated by the Curie tail
($\sim 1/T$) produced by these orphan spins at low temperature. 
Both in $d=2$ and $d=3$, the full susceptibility, when vacancies are placed at 
random, is that of {\it independent} orphans to a good approximation despite
the long-ranged interaction present between them. This persists down to the lowest
temperatures not only because of the  screening of the interactions at
finite physical temperature, and because the size of the Coulomb 
coupling derived from the entropic interaction is comparatively weak,  
but also because the physical orphan spins are related to the degrees
of freedom in the Coulomb antiferromagnet via a sublattice-dependent staggering
transformation, so that the {\it uniform} susceptibility of the physical
orphan-spins corresponds to the staggered susceptibility of the degrees
of freedom of our Coulomb antiferromagnet, and therefore remains largely
unaffected by the fact that the total (vector) gauge charge of our Coulomb
antiferromagnet vanishes.

The remainder of this paper is structured as follows. In Section~\ref{Setting},
we first provide a self-contained review of earlier work on vacancy-induced
effective spins in a class of classical antiferromagnets on lattices consisting
of ``corner sharing units'', and then build on this to provide a careful
derivation of the $T\rightarrow 0$ limit of the emergent entropic interactions
between orphan spins and use this to define our model Coulomb antiferromagnet.
After outlining our analytical and numerical approaches in Sec.~\ref{Methods},
we present the results obtained in $d=2$ and $d=3$ (Sec.~\ref{Results}).
Sec.~\ref{RMT} contains the analysis of the problem in terms of a Euclidean
random matrix while the role of screening and the connection to a scattering
problem are discussed in Sec.~\ref{Screening}. We conclude with a discussion
of these results, and relegate sundry details (such as dicussions of the fully
occupied lattice and the ordered state seeded by a certain finite-lattice
regularization of the two-dimensional Coulomb interaction) to Appendices.  

\section{The random Coulomb antiferromagnetic Hamiltonian}
\label{Setting}

We thus study a classical Heisenberg model
\begin{align}
H = \frac{1}{2}\sum_{i,j} J_{ij} \vec{n}_i\cdot\vec{n}_j .
\label{Hamilton}
\end{align}
where $J_{ij}$ takes on a Coulomb form,
\begin{eqnarray}
J_{ij} &=& -A \log(r_{ij}/\mathcal{L}) \ \ (d=2)\label{Jij2d}\\
         &=& A/r_{ij} \ \ \ \ \  (d=3)\label{Jij3d} .
\end{eqnarray}
This form with $\mathcal{L}$ larger than any $r_{ij}$ has the property that the
interactions are uniformly antiferromagnetic as well as long-ranged.

We need to supplement this by defining the degrees of freedom, unit vectors
$\vec{n}_i$, appearing in Eq.~\ref{Hamilton}.
We concentrate on the case where their locations, denoted by $i$ are chosen
randomly on a square (cubic) lattice in $d=2$ ($d=3$), at a dimensionless
density of $x$ spins per lattice site. 

For long-range interactions like this Coulomb interaction, choices about
boundary conditions or ensemble constraints can be considerably less innocuous
than for short-range systems. In order to illustrate this, and to make natural
choices for  these items, as well as for motivation of our study, we discuss
the derivation of a  random Coulomb antiferromagnet as an effective Hamiltonian
of a diluted Coulomb spin liquid next.

\subsection{Orphan spins and their interactions in diluted Heisenberg antiferromagnets}
\begin{figure}
\epsfig{file=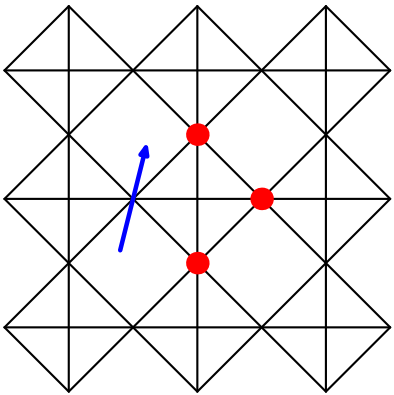,width=0.6\linewidth,angle=0}
\caption{Illustration of the Orphan spin arising from the introduction of
non-magnetic impurities. Its effective moment is half of that of a free spin.}
\label{OrpOnLatt}
\end{figure}
We thus begin by providing a self-contained review of earlier work
on vacancy-induced effective spins in a class of classical frustrated antiferromagnets
on lattices consisting of ``corner sharing units''. The
centers of these in turn define a so-called premedial lattice,
which is bipartite in practically all instances of popularly studied classical
Heisenberg spin liquids \cite{Henley2010coulomb}. A simple model of nearest neighbour
antiferromagnetically interacting spins on such lattices can be written as
\begin{align}
H = \frac{J}{2}\sum_{i, j}\vec{S}_i\cdot\vec{S}_j
  = \frac{J}{2}\sum_{\XBox}(\sum_{\vec{l} \in \XBox}\vec{S}_{\vec{l}})^2 ,
\end{align}
where the summation in the alternate form of the Hamiltonian is carried over
the corner sharing simplices $\XBox$, which might be tetrahedra, as e.g., in
a pyrochlore lattice, triangles in a Kagome lattice, or a combination of both as in
the case of SCGO, and the spins of the frustrated magnet are now labeled by $\vec{l}$,
the links of the bipartite pre-medial graph (whose
sites correspond to the centers of the simplices $\XBox$ of the original lattice,
and links $\vec{l}$ correspond to sites of the original lattice). When written in
this form, it is clear that ground-states are characterized by the constraints:
\begin{align}
\sum_{\vec{l} \in \XBox} \vec{S}_{\vec{l}} = 0 , \forall \XBox ,
\label{eq:vecS}
\end{align}
These local constraints lead to an effective description in terms of a theory of
emergent electric fields that obey a Gauss law.
To see this, we define electric fields
${\mathbf E}^{\alpha}_{\vec{l}} = {\mathbf \epsilon}_{\vec{l}} S^{\alpha}_{\vec{l}}$
on links $\vec{l}$, where ${\mathbf \epsilon}_{\vec{l}}$ is a spatial
unit vector that points from the $A$- to the $B$-sublattice of the premedial lattice
end of this link. The ground-state condition then translates to the statement that
the lattice-divergence of this electric field vanishes at each site $\XBox$ for each $\alpha$. 
The key idea of this effective description is that the coarse-grained (entropic) free energy
density depends quadratically on the local electric field, and deviations from the vanishing
divergence condition amount to the appearance of vector Coulomb charges~\cite{sen2012vacancy}.
These emergent gauge charges are defined for each lattice point $\XBox$ of the
bipartite premedial lattice:
\begin{align}
\vec{Q}_{\XBox} = \eta(\XBox) \sum_{\vec{l} \in \XBox} \vec{S}_{\vec{l}} ,
\end{align}
and the staggering factor, $\eta(\XBox) = +1$ if $\XBox$ is an $A$-sublattice site
of the premedial graph and $-1$ otherwise.
Since each microscopic spin contributes with opposite signs to the vector charge
on two neighbouring simplices, the total gauge charge of a system without
boundaries  must vanish in every configuration of the system
\begin{align}
\sum_{\XBox}\vec{Q}_{\XBox} = 0 .
\label{TotChargeVan}
\end{align}
This very natural condition--akin to the charge-neutrality of the full universe, and in our case unavoidable
due to the microscopic origin of the emergent gauge charge -- will be explicitly
imposed in our Monte Carlo simulations of the system.

The mapping of the pure system to an emerging gauge field theory at low
temperatures makes clear that generalized ``vector charges'', $\vec{Q}_{\XBox}$,
are generated thermally as a consequence of the violation of the ground
state constraints. The constraint Eq. \ref{eq:vecS} is also unavoidably violated in the
presence of non-magnetic impurities (Fig.~\ref{OrpOnLatt}) whenever all but
one spin of a given simplex are substituted for by vacancies (simplices containing
at least two spins can in general satisfy the zero total spin condition and such
simplices do not host a vector charge in the $T\rightarrow 0$ limit). 
Indeed, when all spins but one in a simplex are replaced by vacancies, the
result is a paramagnetic Curie-like response
\cite{moessner1999magnetic,SenDamleMoessnerPRL,sen2012vacancy}, which dominates the
susceptibility response at low temperatures. The lone spins on these
defective simplices, which serve as the epicenter of this paramagnetic response,
were baptized {\it Orphans} (Ref. \onlinecite{schiffer1997two}) in the first
studies of this effect.

The field theory developed in Refs.~\onlinecite{SenDamleMoessnerPRL,sen2012vacancy} extends
the self consistent gaussian approximation (SCGA)~\cite{garanin1999classical},
a theory successful in describing low temperature correlations on the undiluted
systems, to incorporate the effects of dilution and study the physics of these
orphan spins at non-zero temperature in a manner that treats entropic effects
on an equal footing with energetic considerations. In its original form the
SCGA replaces the hard constraint on the spins norm, $\vec{S}_i^2 = S^2$,
by the relaxed {\it soft spin} condition on their thermal average $\la\vec{S}_i^2\ra = S^2$.
The key insight of Refs.~\onlinecite{SenDamleMoessnerPRL,sen2012vacancy}, that
led to the detailed analytical understanding summarized below, was the
following: While
it is sufficient to treat in this self-consistent Gaussian manner all spins
other than the lone orphan spin in a simplex in which all but one spin has
been replaced by vacancies, this is much too crude an approximation
for the orphan-spin itself, which must be treated without approximation
as a hard-spin obeying $\vec{S}_{\rm orphan}^2 = S^2$. Remarkably, the resulting
hybrid field theory continues to be analytically tractable when the number
of orphan spins is small\cite{SenDamleMoessnerPRL,sen2012vacancy}.
With just one orphan present in a sample with an external magnetic field of
strength $B$ along the $z$ axis, the theory predicts that this orphan spin sees a magnetic
field $B/2$, with the other half of the external field screened out by the coupling
to the bulk spin-liquid. The resulting polarization of the orphan serves as a source
for an oscillating texture that spreads through the bulk. The net spin carried
by the texture cancels half the spin polarization of the orphan-spin, resulting
in an impurity susceptibility corresponding to a classical spin $S/2$.
With more than one orphan present, the spin-textures seeded by each orphan
mediate an effective entropic interaction between each pair of orphan spins.

The effective action for a pair of orphans is predicted in this manner to have
the form
\begin{align}
-\beta J_{\text{eff}}(\vec{r},T)\vec{n}_1\cdot\vec{n}_2 ,
\end{align}
where $\vec{n}$ are unit-vectors corresponding to the directions of the orphan-spins
in a given configuration.
The exchange coupling has a particularly simple form in the large separation limit
\begin{align}
\beta J_{\text{eff}} \approx -\eta(\vec{r}_1)\eta(\vec{r}_2)\frac{\la\vec{Q}_{\XBox}(\vec{r}_1)\cdot\vec{Q}_{\XBox}(\vec{r}_2)\ra}{\la\vec{Q}_{\XBox}\cdot\vec{Q}_{\XBox}\ra^2}
\label{JEffAsChChCorr}
\end{align}
which involves only ``charge-charge'' correlations {\it calculated in the
pure system}:
\begin{align}
\la\vec{Q}_{\XBox}(\vec{r}_1)\cdot\vec{Q}_{\XBox}(\vec{r}_2)\ra \sim &-T^2 T^{d/2-1} \nonumber\\
&\times\int^{\Lambda/\sqrt{T}}d^{d}q\frac{\exp(i\vec{q}\cdot(\vec{r}_1-\vec{r}_2))}{\Delta_c q^2 + \kappa}.
\label{eq:07}
\end{align}
The denominator of Eq.~\eqref{JEffAsChChCorr}, behaves at low temperatures as
$\la\vec{Q}_{\XBox}\cdot\vec{Q}_{\XBox}\ra = T/J$ from equipartition.

For orphans in $d=2$, one finds:
\begin{align}
J_{\text{eff}}(\vec{r}_1 - \vec{r}_2,T) = \eta(\vec{r}_1)\eta(\vec{r}_2)T\mathcal{J}(|\vec{r}_1 - \vec{r}_2|/\xi_{\text{ent}}) 
\end{align}
with an entropic screening length $\xi_{\text{ent}} = 1/\kappa \sim 1/\sqrt{T}$
separating two regimes for ${\cal J}(\kappa r)$. For $\kappa r\ll1$ a
logarithmic one, ${\cal J}(\kappa r)\sim -\log(\kappa r)$; and for 
 $\kappa r\gg1$ a screened regime, ${\cal J}(\kappa r)\sim \frac{1}{\sqrt{\kappa r}}\exp(-\kappa r)$.
Analogously in $d=3$, 
\begin{align}
J_{\text{eff}}(\vec{r}_1 - \vec{r}_2,T) = \eta(\vec{r}_1)\eta(\vec{r}_2)T^{3/2}\mathcal{K}(|\vec{r}_1 - \vec{r}_2|/\xi_{\text{ent}}) 
\end{align}
the  entropic screening length $\xi_{\text{ent}} = 1/\kappa \sim 1/\sqrt{T}$ separates
 two regimes, algebraic $\mathcal{K}(r)\sim -1/r$ and screened
$\mathcal{K}(r)\sim \exp(-\kappa r)$.

In the physical system, at any nonzero temperature, this is thus a `short-ranged' interaction on account of the 
finite screening length which, however, 
diverges as $1/\sqrt{T}$. In this article, we are interested in the limit
of $T=0$, where the interaction takes on the novel -- for magnetic systems -- long-range Coulomb form. 

\subsection{Model Hamiltonian}
In the limit of $T \to 0$, we are thus led by these considerations to 
Coulomb interactions between the vector orphan spins, which we here study in detail. 
For simplicity, we consider unit-vector spins $\vec{n}$ at random locations in
a periodic hypercubic lattice of linear size $L$ with occupancy probability $x$,
corresponding to an underlying spin liquid on the checkerboard and ``octochlore''
lattices of corner-sharing units involving $2^d$ spins in $d$ dimension.

In what follows, we will get rid of the sublattice factors that affect the sign
of the effective interaction by inverting all unit-vectors placed on the $B$ sublattice.
In other words, we identify $S \vec{n}_i$ with $\eta_i \vec{S}_{{\rm orphan}, i}$, where
$\vec{S}_{{\rm orphan}, i}$ is the orphan spin on the simplex labeled by $i$ in
the underlying diluted frustrated magnet.

This gives us a ``random Coulomb antiferromagnet'' in which unit-vector spins
interact with an exchange coupling that is always antiferromagnetic but of a
long-range Coulomb form at large distances.
For a classical system, this transformation is innocuous, but note that natural
observables cease to be so under this mapping -- e.g. the orphan spin contribution
to the uniform susceptibility of the underlying diluted magnet is now given by the
staggered  susceptibility of our Coulomb antiferromagnet. 

As is usual for entropic interactions in the limit of $T \to 0$, the strength of
their coupling, $A$, is fixed by the microscopics of the model from which they
have emerged. In this work, we are interested in exploring the generic behaviour
of such models -- in particular, identify possible phases -- and thus allow the
coupling $A$ to be variable. For completeness, we mention that $A=\frac{1}{4\pi}$
for the checkerboard lattice.

This therefore leads to the form of $H$ at the beginning of this section,
Eq.~\ref{Hamilton}. To make Eq.~\ref{Jij2d} dimensionally unambiguous we write:
\begin{align}
J_{ij} = -A \log(r_{ij}/{\mathcal L}) \ \ (d=2) \nonumber
\end{align}
with ${\mathcal L}$ conveniently set to a value of order the system size $L$
so that $J_{ij}>0$ always.
In the above language, with sublattice factors $\eta$ absorbed into
the definitions of $\vec{n}_i$, the zero gauge-charge constraint imposed by the
microscopic origin of this effective model now translates to the constraint that
$\sum_{i}\vec{n}_{i} = 0$ in every allowed configuration of our Coulomb
antiferromagnet. This constraint in fact can also be imposed by adding an infinitely
strong interaction acting equally between all spins. This equivalence renders the
detailed choice of $\mathcal{L}$ immaterial.

We note an interesting scale-invariance of this model in the limit of small
densities of spins. This scale invariance is inherited from that of the logarithmic
function under scaling transformations: $J(\kappa r) = \log(\kappa) + J(r)$,
together with net charge neutrality Eq.~\eqref{TotChargeVan}:
\begin{align}
\sum_i \vec{n}_i = 0 ,
\label{EffTotChargeVan}
\end{align}
implies that the extra term $\log(\kappa)$ gives a temperature-independent 
contribution to the action determined by 
$1/2\sum_{i \neq j} \vec{n}_i\cdot\vec{n}_j = -N/2$.  The partition function thus only 
picks up a constant factor:
\begin{align}
Z' = e^{-\beta\sum_{i,j}J(\kappa r_{ij})\vec{n}_i\cdot\vec{n}_j} =  e^{\beta\log(\kappa)N/2}Z ,
\end{align}
It also means that, rather unusually, in the continuum limit  $x\rightarrow0$
the partition function is a scaling function depending on the randomly chosen orphan locations
only scaled by their mean separation. Lattice discretisation effects at finite $x$ 
break this equivalence.
The scaling transformation for the model in three dimensions gives $J(\kappa r) = J(r)/\kappa$,
what implies for the partition function a rescaling of $\beta$:
\begin{align}
Z'(\beta) = e^{-\beta\sum_{i,j}J(\kappa r_{ij})\vec{n}_i\cdot\vec{n}_j} =  Z(\beta/\kappa) .
\end{align}

For Coulomb interactions in a finite-size system, various choices of the interaction
yield the same large-distance form in the limit $L \to \infty$. The most natural
form from the point of view of the effective field theory predictions for
emergent interactions between orphan spins is the Fourier transform of the
inverse of the lattice Laplacian,
$d - \sum\limits^d_{i = 1} \cos k_i$:
\begin{align}
\label{LGFDef}
J(r_{ij}) = \frac{\pi}{L^2}\sum_{\vec{q}}\frac{e^{i\vec{k}\cdot\vec{r}_{ij}}}{d - \sum\limits^d_{i = 1} \cos k_i}.
\end{align}
This we call the lattice Green function (LGF), and our most detailed studies
are carried out with this form of the interaction.

Alternatively, one can work directly with the Coulomb form, e.g. for $d = 2$:
\begin{align}
\label{LOGDef}
J(r_{ij}) = -\log\left(\frac{r_{ij}}{{\mathcal L}}\right).
\end{align}
with ${\mathcal L} =L/\sqrt{2}$.
\begin{figure}
\epsfig{file=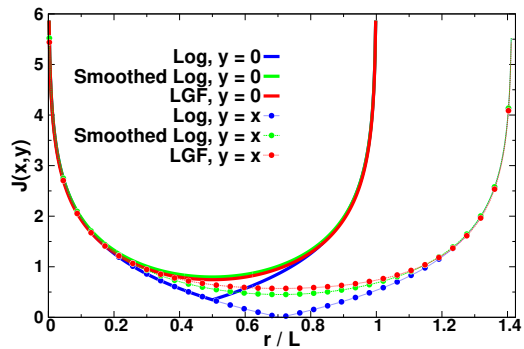,width=0.8\linewidth,angle=0}
\caption{$J(x,y)$ used in the simulations in $d=2$.}
\label{Int}
\end{figure}
This form agrees with the LGF interactions at short distances (see Fig.~\ref{Int}).

The issue of how to impose the boundary conditions, and therefore how to compute
$r_{ij}$, turns out to make much difference on the results for a finite system, as we shall see. The choices
of either 
\begin{align}
r_{ij} = |\vec{r}_i - \vec{r}_j| = \sqrt{\tilde{x}_{ij}^2 + \tilde{y}_{ij}^2} ,
\label{AnisotDist}
\end{align}
with $\tilde{x}_{ij} = \text{min}(|x_i -x_j|, L - |x_i -x_j|)$ 
or
\begin{align}
r_{ij} = \frac{L}{\pi}
        \sqrt{\sin^2\left(\frac{\pi(x_i - x_j)}{L}\right) +
              \sin^2\left(\frac{\pi(y_i - y_j)}{L}\right)} ,
\label{IsotDist}
\end{align}
result in different behavior for the system, which will be explained in
more detail in the results section. We refer to these choices as periodised,
and smoothed, logarithms, respectively. The latter is very close to the LGF,
while the former maintains a finite difference to it at the periodic boundary,
where it is not differentiable for any $L$ (Fig. \ref{Int}).
It is easily seen why this finite difference is independent of $L$, if one
compares the smoothed log to the periodized Log, approximatelly equivalent to
comparing the LGF with the Log. Looking, e.g., at the midpoint of one edge
($x_{ij}=L/2$, $y_{ij}=0$) one finds:
\begin{align}
(J^{\text{LGF}}_L - J^{\text{Log}}_L)(L/2,0) \approx\log(\pi/2),
\end{align}
where the subindex $L$ emphasizes that we are looking at the respective forms
of the interactions in a finite system of size $L$.

Note, again, that adding a constant to the interaction (in $d = 2$), e.g., by changing the
denominator of Eq.~\eqref{LOGDef}, leaves the interaction unchanged due to the
global charge neutrality constraint.

\section{Methods}
\label{Methods}
The analysis of spin systems with the potential for glassy phases is a
delicate endeavour as equilibration of large systems is elusive. Existence and
determination of a transition temperature is usually a
controversial issue\cite{lee2007large, pixley2008large}. Since our system has
long ranged interactions, boundary effects can cause yet more trouble.
This is why we combine analytical with numerical methods, as well as mappings to
other problems which have received attention in a different context previously.

Numerically, we study the behaviour of this model through Monte Carlo (MC)
simulations, and analytically in the self-consistent Gaussian (``large-m'',
also denoted in the following as LM approach~\cite{hastings2000ground,aspelmeier2004generalized,lee2005spin}) approximation, where the
parameter $A$ mimics an inverse temperature. 
Our MC simulations directly impose the constraint, Eq.~\eqref{TotChargeVan}. For that we
initialize the system in a random configuration of vanishing total spin, and the
update movements on the system consist of selecting an arbitrary pair of spins, and
rotating them around the axis determined by their vectorial sum. A MC simulation of
the same system with strictly positive interactions, without this constraint on the
total spin has been also investigated, and the conclusion is that while the relaxation
time increases, the system still prefers to stay close to the manifold of vanishing
total spin.

The LM approach consists of considering spins with $m$ components and letting
$m\rightarrow\infty$. This is formally equivalent to the soft spin approximation and it
only gives in principle information about the infinite number of components limit, but
this can be understood as the 1st term in an expansion of the $O(m)$ model.
It has been very successful in the analytical study of correlations in highly
frustrated spin systems~\cite{garanin1999classical}, being able to reproduce the
main features of the on-going phenomena, such as existence of long range dipolar
correlations at $T=0$, characterized by the presence of ``pinch points'' in the
structure factor~\cite{isakov2004dipolar}.

The LM approach allows an analysis of the system both at finite coupling
strengths $A<\infty$, and at $A=\infty$. The study of glassiness with this approach
has been already undertaken in a variety of models~\cite{lee2005spin,beyer2012one},
and we will be following a similar methodology. Correlations are computed through
the matrix:
\begin{align}
B_{ij} = J_{ij} + h_i \delta_{ij} ,
\end{align}
and are given by:
\begin{align}
C_{ij}=\frac{1}{m}\la\vec{n}_i\cdot\vec{n}_j\ra = \frac{1}{A}(B^{-1})_{ij} .
\end{align}
These can be computed once the Lagrange multipliers, $h_i$, are determined
through the set of nonlinear equations:
\begin{align}
C_{ii} = 1 .
\end{align}
For comparison between LM and MC, we scale observables and couplings with
$m$ so that their small-coupling (``high-temperature'') forms agree.

The point $A=\infty$ is treated within the LM approach by determining the
(unique~\cite{hastings2000ground}) ground state through a {\it local field quench}
algorithm~\cite{walker1980computer}. This algorithm is based on the fact that if the number of spin
components, $m$, is large enough (larger than $\sqrt{2N}$\cite{hastings2000ground}),
then a system of spins with $m$ components is effectively equivalent
to the corresponding system in the limit $m\rightarrow\infty$.
The algorithm then consists of taking a system of $N$ spins with $m>\sqrt{2N}$
components initially randomly oriented, and then iteratively aligning each spin
with its local field. This procedure is expected to converge to the unique
ground state, from which all the quantities of interest can be computed.

A fundamental quantity at $A=\infty$ within the LM approach is the number of zero eigenvalues,
$m_0$, of the matrix $B_{ij} = J_{ij}+h_i\delta_{ij}$; it can be shown\cite{hastings2000ground}
that the ground state spin vectors span an $m_0$ dimensional space. This
quantity should scale with the number of particles in the system as $m_0\sim N^\mu$.
Furthermore, as was shown in Ref.~\onlinecite{lee2005spin}, the same exponent
controls the scaling of the spin glass susceptibility for the ground state configuration:
$\chi_{SG}\sim N^{1-\mu}$.

The main quantity of interest in our study will be the spin glass susceptibility
(square brackets here and throughout indicate the disorder average),
\begin{align}
\chi_{SG}(\vec{k}) = \left[\frac{1}{N}\sum_{i,j}\la\vec{n}_i\cdot\vec{n}_j\ra^2
                                  \cos{\vec{k}\cdot(\vec{r}_i-\vec{r}_j)}\right] ,
\end{align}
obtained in the MC simulations through the overlap tensor~\cite{binder1986spin}:
\begin{align}
Q^{\alpha ,\beta}_{\vec{k}} = \frac{1}{N}\sum_i{n^\alpha_{i,1}n^\beta_{i,2}}e^{i\vec{k}\cdot\vec{r}_i},
\end{align}
where greek indices refer to the spin components, while the indices $1,2$ refer to
two independent replicas of a disorder realisation. This might be interpreted as
the overlap of a spin configuration with itself after an infinitely
long time. Since the onset of glassiness can be also understood as a divergence of
the equilibration time, the nonvanishing of this order parameter signalizes the
transition. 

The spin glass susceptibility in terms of this tensor is:
\begin{align}
\chi_{SG}(\vec{k}) = \left[N\sum_{\alpha , \beta}{\left\la\left|Q^{\alpha , \beta}_{\vec{k}}\right|^2\right\ra}\right].
\end{align}
We follow the usual practice to determine the spin glass transition by computing
a finite system correlation length associated to the susceptibility above. The
Ornstein-Zernike form for correlations gives:
\begin{align}
\xi_L = \frac{1}{2\sin(k_{min}/2)}\left(\frac{\chi_{SG}(0)}{\chi_{SG}(\vec{k}_{min})}-1\right)^{1/2} ,
\end{align}
and near the transition, the finite size scaling prediction is expected to be:
\begin{align}
       \frac{\xi_L}{L}=X(L^{1/\nu}(1/A-1/A_{c})) ,
\end{align}
while the susceptibility should follow:
\begin{align}
       \frac{\chi_{SG}}{L^{\gamma/\nu}}=Y(L^{1/\nu}(1/A-1/A_{c})) ,
\end{align}
\begin{figure}
\epsfig{file=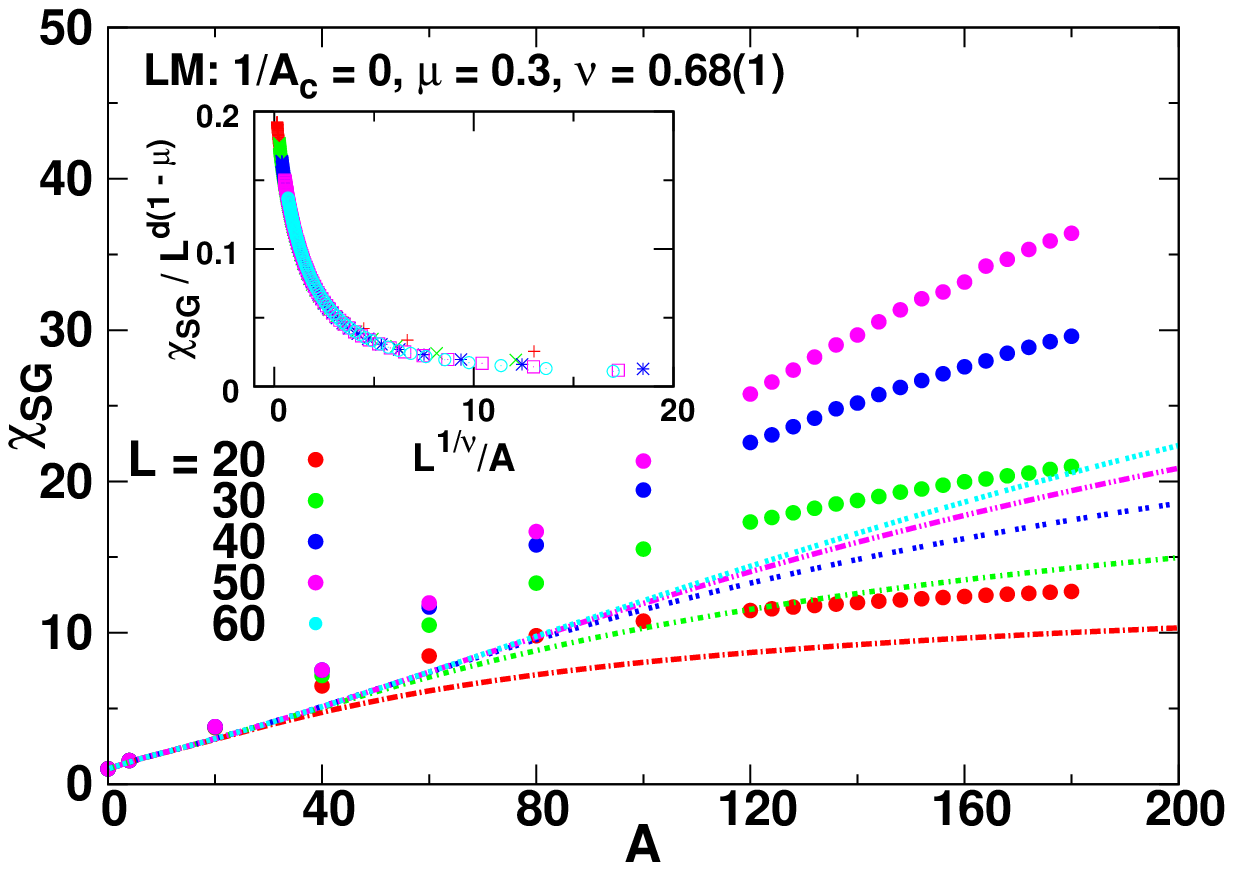,width=0.95\linewidth,angle=0} \\
\epsfig{file=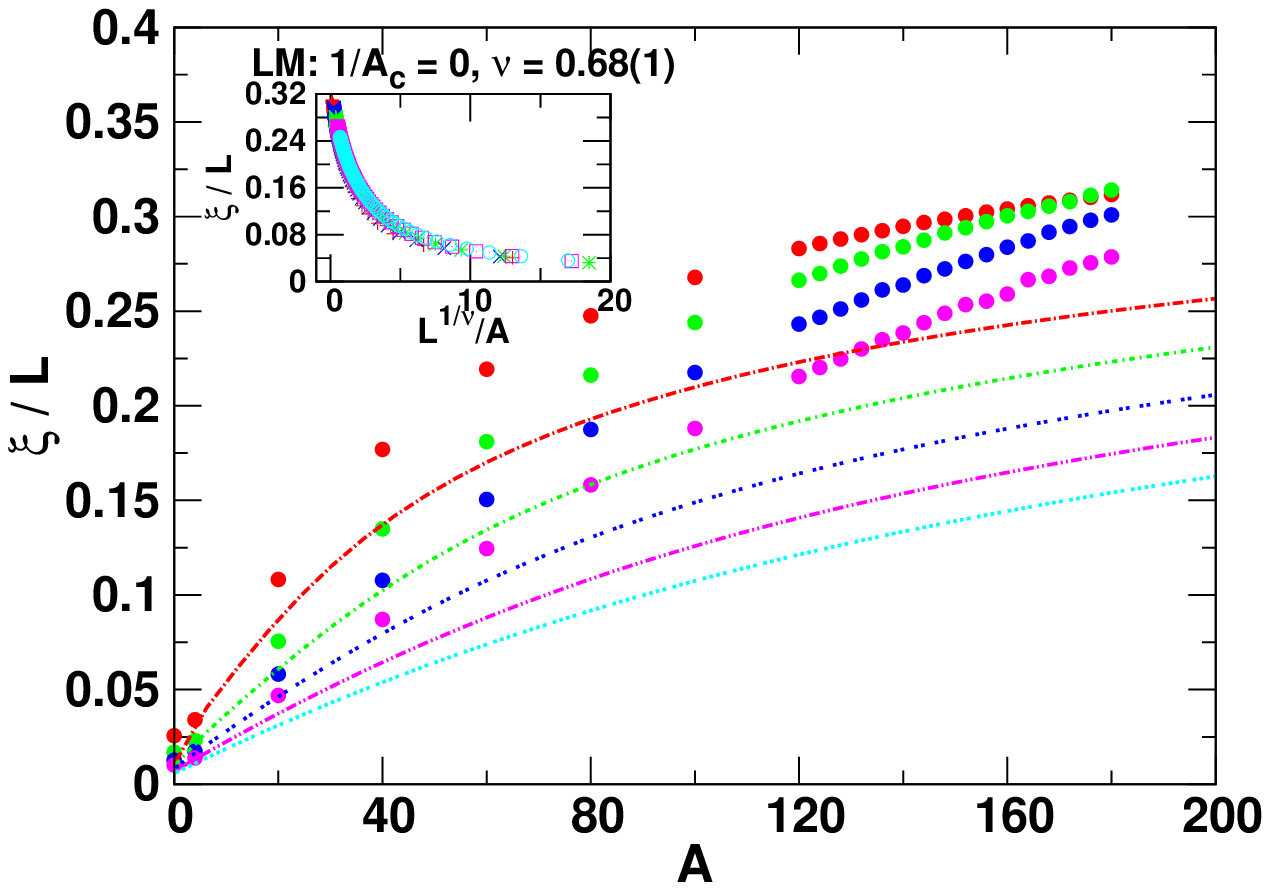,width=0.95\linewidth,angle=0}
\caption{Spin glass susceptibility (top) and correlation length (bottom) for the LGF interaction.
Several system sizes are indicated by different colours. Circles indicate (error bars are of the
order of the circles size) MC simulations, while lines are from the LM approach--correlations
are stronger for Heisenberg spins than the 'soft' LM spins throughout. The insets
show scaling collapse for LM for $1/A_c=0$.}
\label{LMandMCResultLgf}
\end{figure}

Notice that these scaling relations only hold if there exists a crossing of
finite size correlation length curves for different system sizes at an unique
finite coupling strength value. The absence of such a crossing at a finite $A_c$
indicates the absence of a phase transition. Nonetheless a phase transition at
$A_c=\infty$ cannot thus be ruled out and the LM approach allows an analysis in
this situation. The scaling relations predicted to hold in this case ($A_c \rightarrow \infty$) are:
\begin{align}
\chi_{SG} = L^{d(1-\mu)}Y(L^{1/\nu}/A)   , \hspace{0.5cm} \xi_L/L = X(L^{1/\nu}/A) .
\label{AinftyScal}
\end{align}
The exponent $\mu$ here is the  one previously introduced for
the scaling of the number of zero eigenvalues of the matrix $B$ with the number
of particles in the system.


\section{Results}
\label{Results}
\subsection{Two dimensions}

\begin{figure}
\epsfig{file=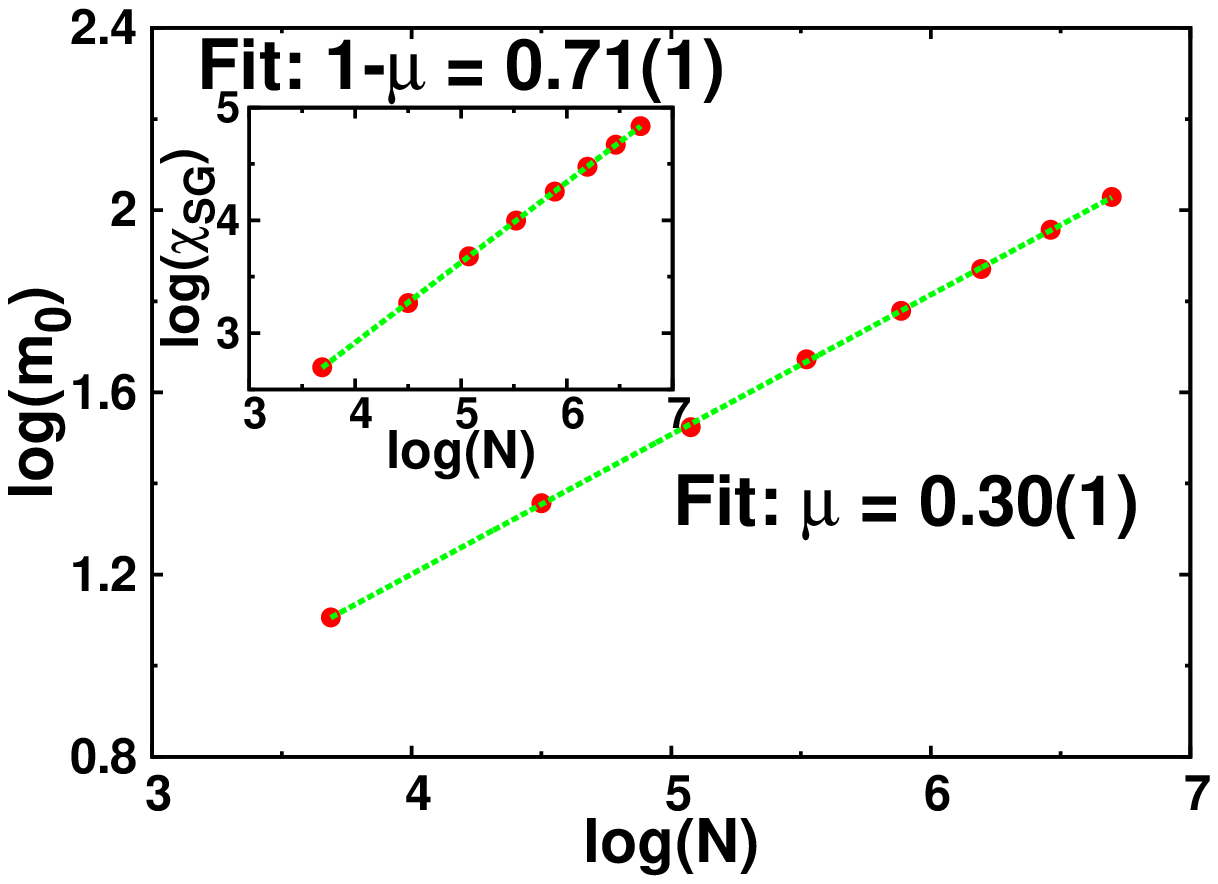,width=0.95\linewidth,angle=0} \\
\epsfig{file=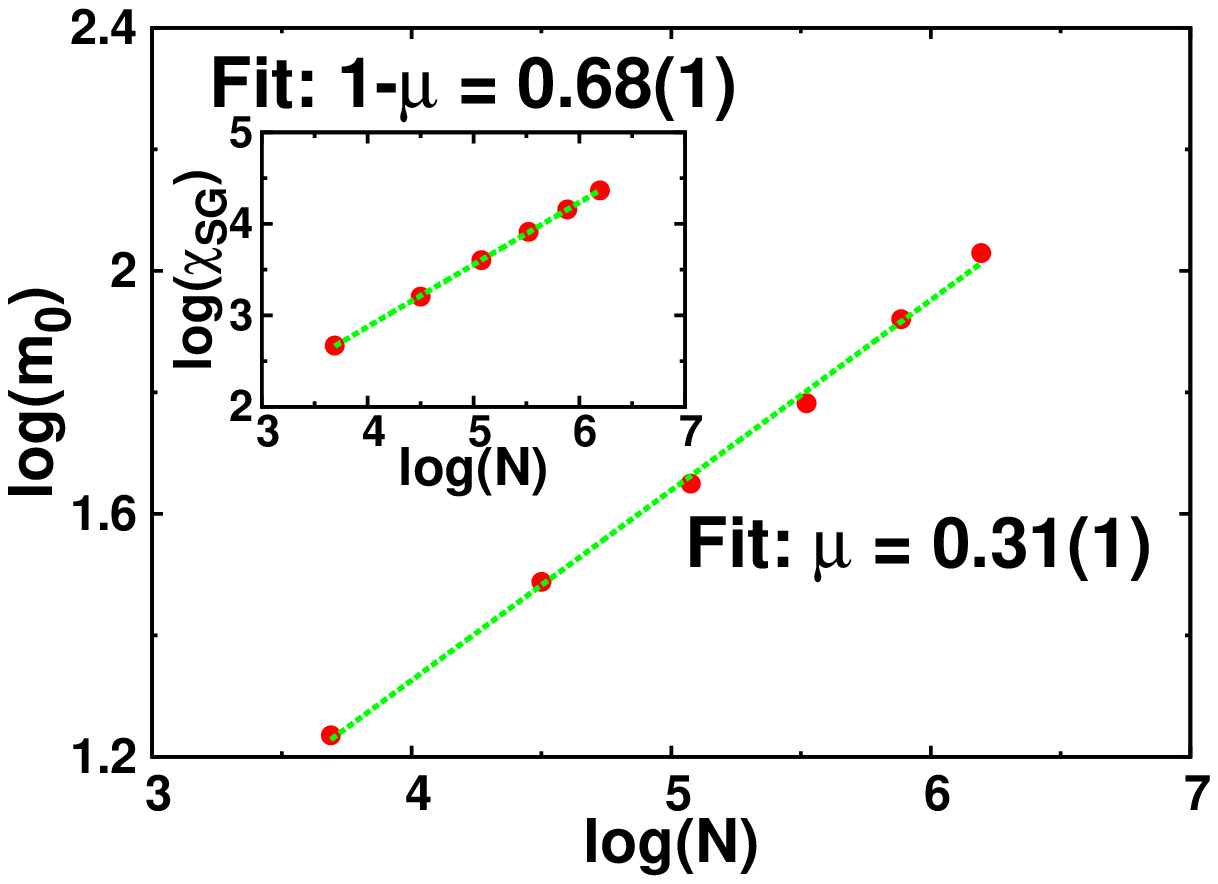,width=0.95\linewidth,angle=0}
\caption{Scaling of the number of zero eigenvalues ($m_0$) of the matrix $B$
defined in the text and of the spin glass susceptibility (insets) with the number of
particles for the LGF (top), and Log (bottom) interactions.}
\label{LMmuScal}
\end{figure}

The two approaches (MC and LM)  yield a broadly consistent picture for each of
the interactions studied. We conduct an analysis of a possible freezing transition
in the model by measuring the spin glass susceptibility 
and trying to identify the transition through a finite size scaling of
its associated correlation length. Other observables such as the specific heat or the
uniform susceptibility were also studied, though these do not indicate any of the
conventional orderings.

The results from MC simulations and LM calculations are shown on Fig.~\ref{LMandMCResultLgf}
for the system with LGF as interaction for a fixed density $x=0.10$
of particles. In each case the number of disorder realisations simulated was $200$.

Globally, correlations are stronger for the  MC simulations on Heisenberg spins compared to the
LM results. This is in keeping with the general lore that a lower number of spin components is
conducive to spin freezing, as is well known from the comparison of Ising and Heisenberg spins. 

In the broad range of coupling strengths considered by our analysis, no unique crossing
for the different system sizes of the correlation length curves can be identified.

The LM analysis at $A=\infty$ yields the exponent $\mu$ as indicated in
Fig.~\ref{LMmuScal}. This seems to have the same value, $\mu\approx0.3$
for both the LGF and Log interactions.

The exponent value $\mu = 0.3$ is used as input, together with the assumption that
$A_c=\infty$ for the LGF, in attempting a scaling collapse of the LM data. The exponent
$\nu$ was determined by a fitting procedure with the scaling relation, Eq. \eqref{AinftyScal},
{\it only using data for the correlation length}.
The resulting scaling collapse is shown on the inset of the lower panel of
Fig.~\ref{LMandMCResultLgf}, where $\nu=0.68(1)$ is obtained. Finally, we use all these
exponents on the predicted scaling relation for the susceptibility (the result is shown on the
inset of the upper panel of Fig.~\ref{LMandMCResultLgf}).
The available data from the LM calculations indicates therefore a freezing transition
at $A_c=\infty$ for the diluted model with LGF as interaction in two dimensions.

\begin{figure}
\epsfig{file=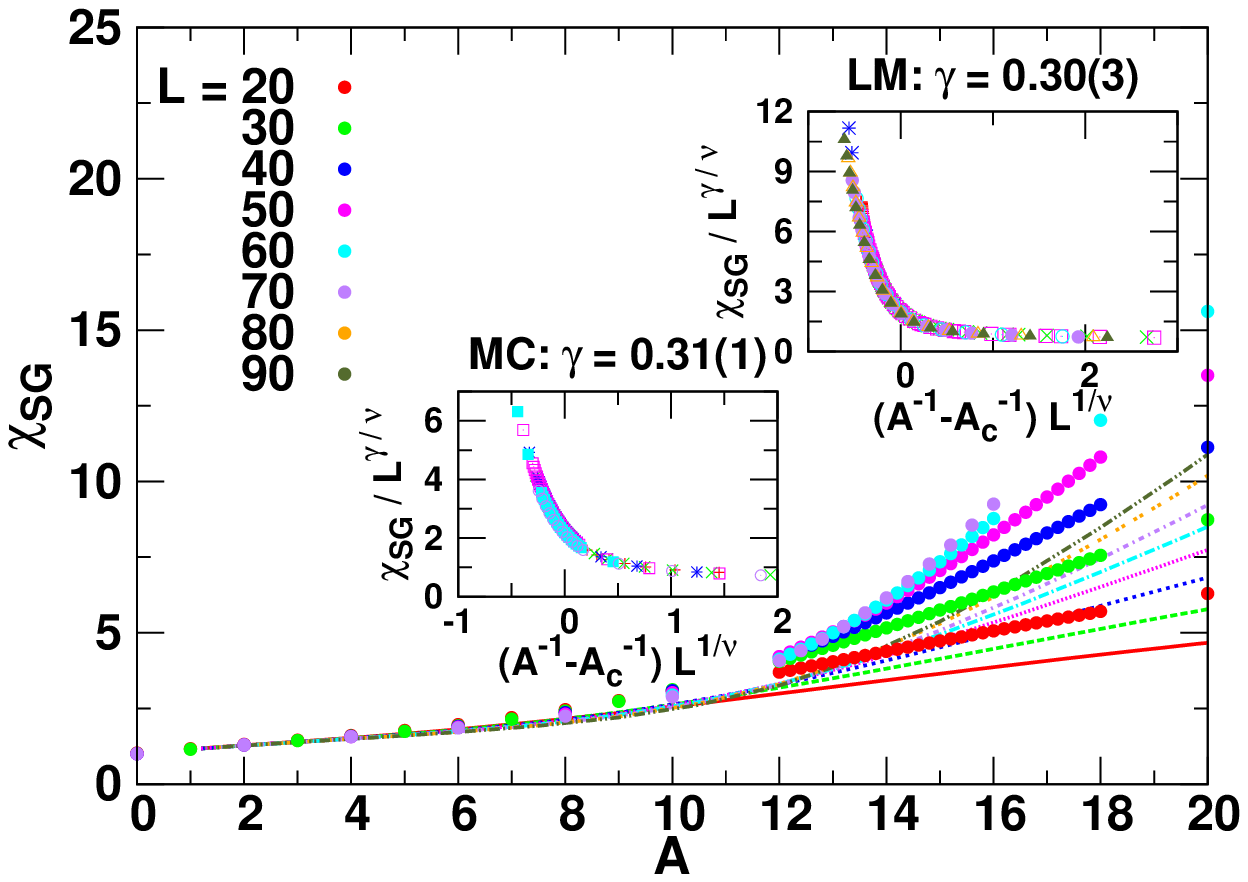,width=0.95\linewidth,angle=0} \\
\epsfig{file=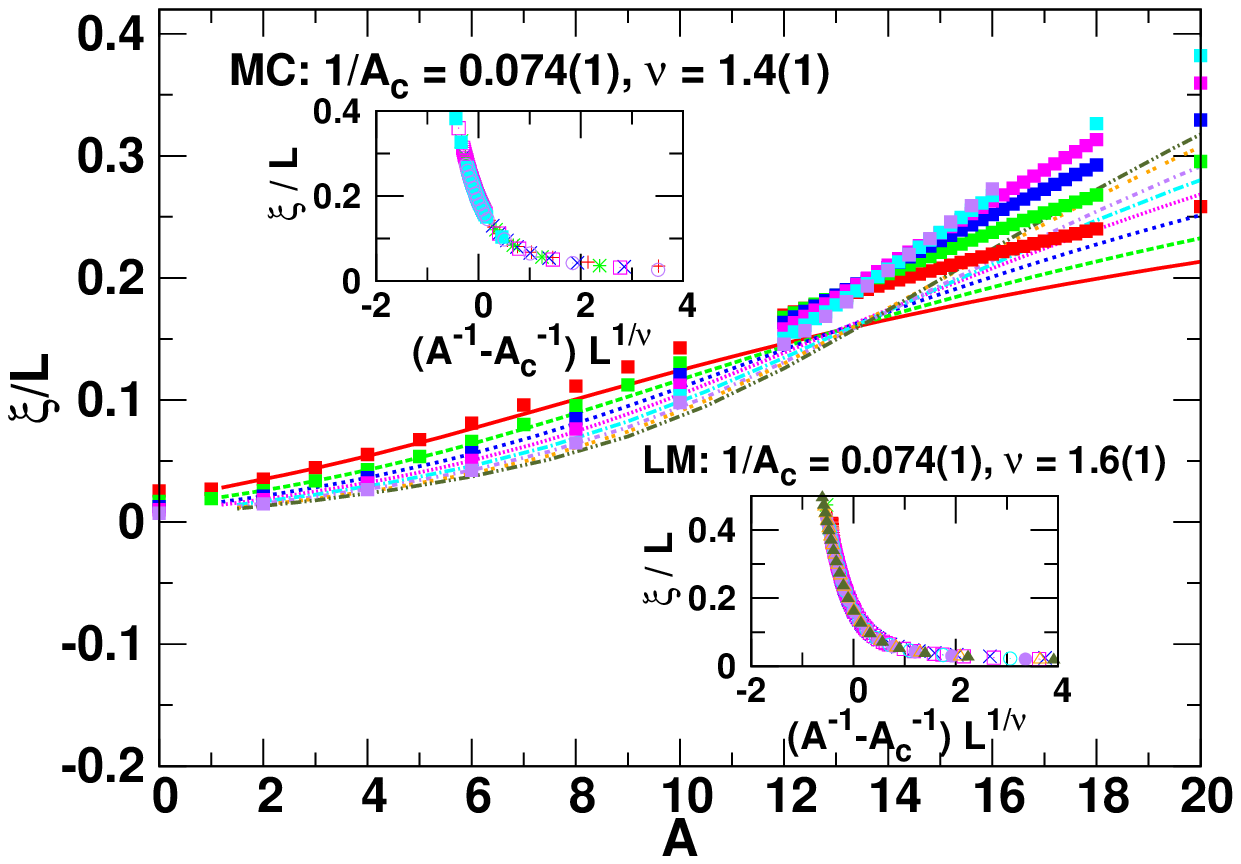,width=0.95\linewidth,angle=0} 
\caption{Spin glass susceptibility (top) and correlation length (bottom) for the Log interaction,
as computed on the MC simulations (points) or with the LM approach (lines). The insets show the
corresponding scaling collapses.}
\label{MCandLMResultLog}
\end{figure}

The Log interaction turns out leads to a dramatically differing behaviour! This is a
surprising result, as the interactions only differ appreciably at large distances
(Fig.~\ref{Int}). Fig.~\ref{MCandLMResultLog} shows the results
for the observables of interest as obtained from MC simulations and LM calculations,
respectively. Here again we fix the density of particles $x=0.1$, and consider
$200$ disorder realisations.
A clear crossing of the correlation length curves for different system sizes occurs
and scaling collapses of the data are possible, which are shown together with the
corresponding critical exponents as insets.

\begin{figure}
\epsfig{file=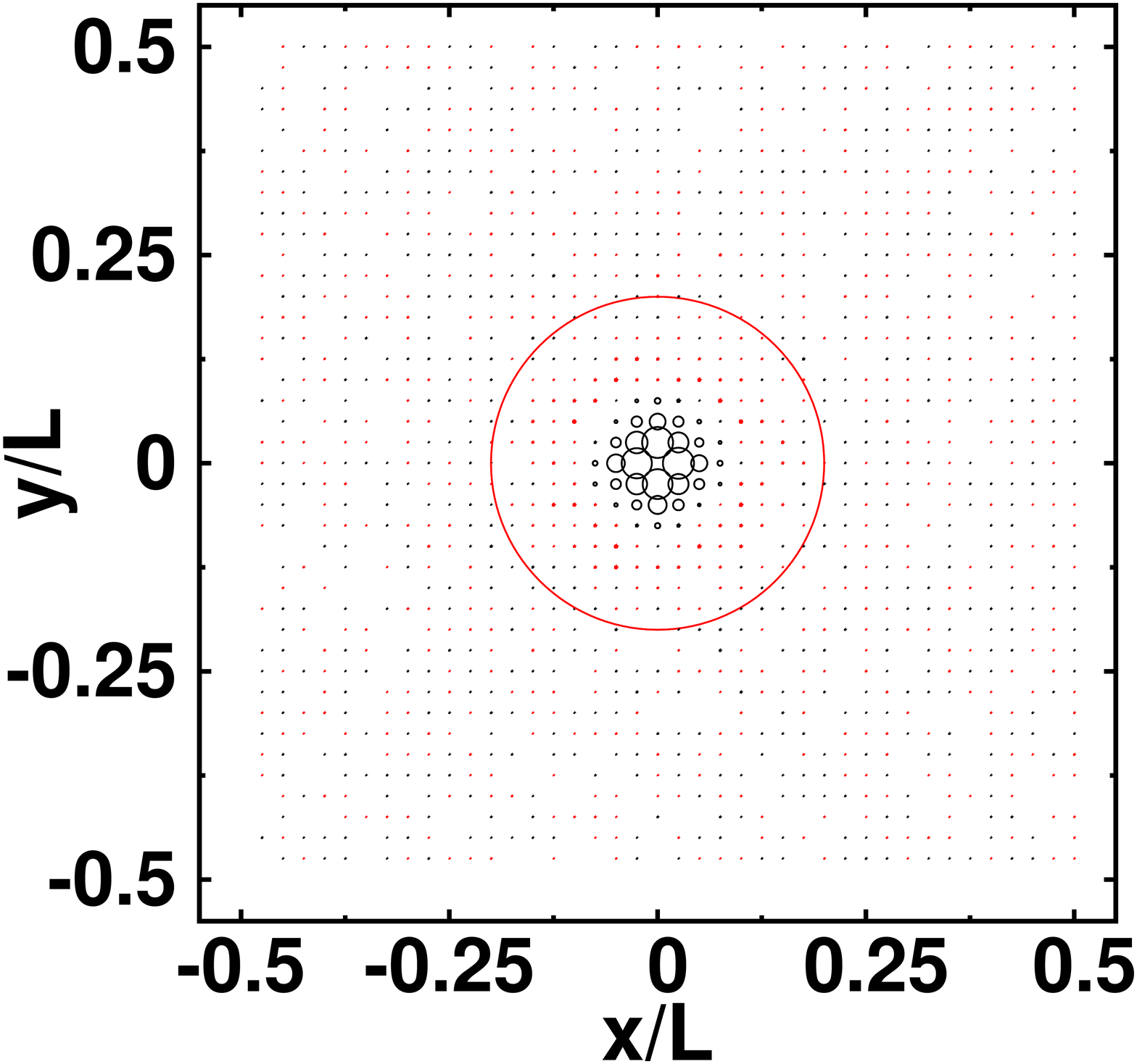,width=0.95\linewidth,angle=0} \\
\epsfig{file=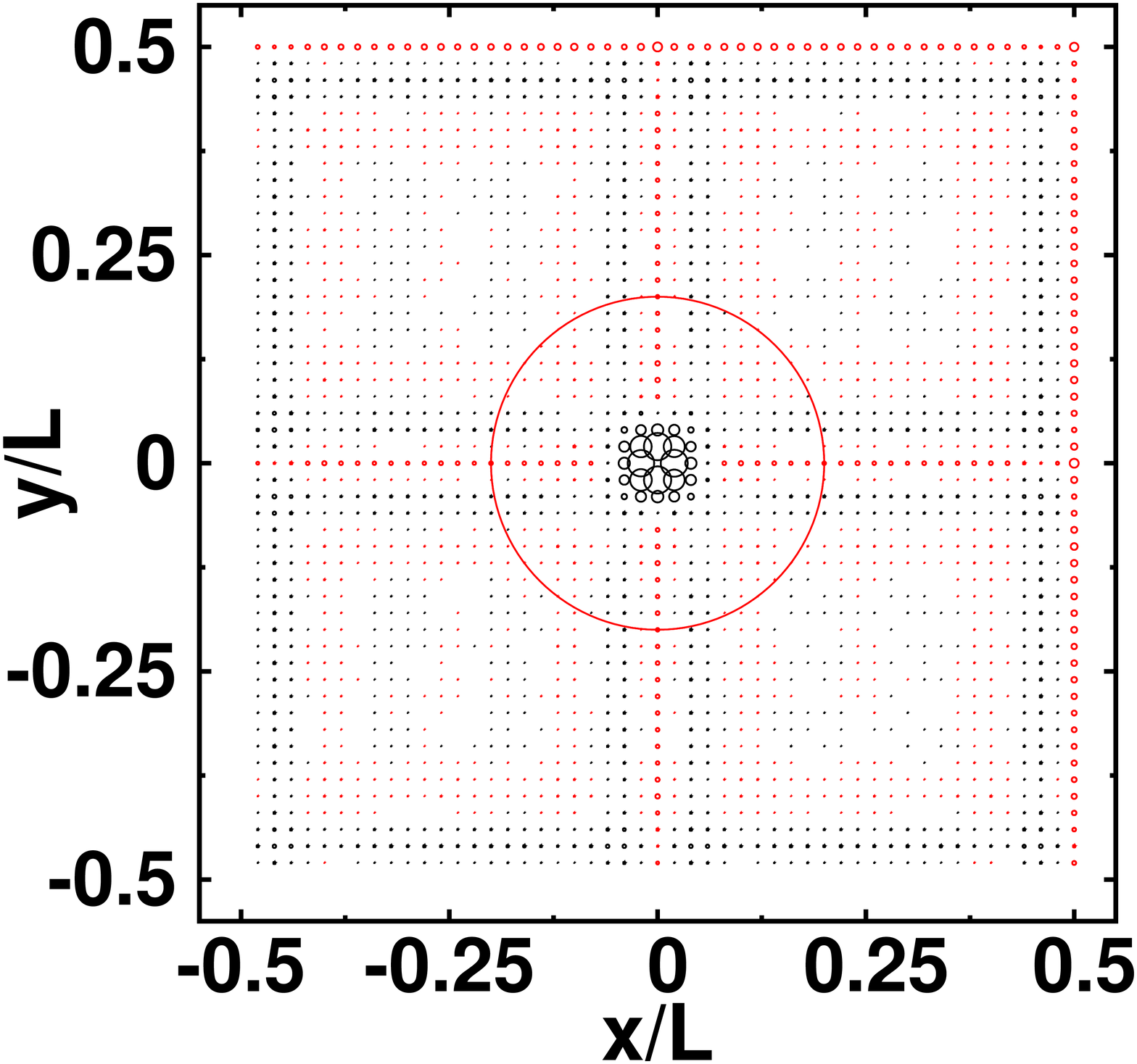,width=0.95\linewidth,angle=0}
\caption{
Disorder-averaged pair correlations with a spin at the origin as a function of
relative coordinates, centers of circles indicate the position of the spin, its
radius gives the magnitude, with red (black) denoting positive (negative) correlations.
The central red circle thus reflects $\la\vec{S}_i^2\ra=1$. The upper panel indicates
the result for the LGF with $A=100$, while the lower corresponds to the Log with $A=20$.
Data shown from MC is in agreement with LM (not shown).}
\label{PCvsXYfromMC}
\end{figure}

To study more closely this effect, we consider the pair correlations as a function of
the relative coordinates of the pairs, averaged over disorder realisations (Fig. \ref{PCvsXYfromMC}).
The profile is isotropic for the LGF with only the 1st few nearest neighbors significantly
antiferromagnetically correlated. On the other hand, the Log interaction yields strongly
anisotropic behavior (the interaction itself is anisotropic) and this seems to be
responsible for what we see as a ``glassy phase transition'' emerging from the
``splaying out'' of the susceptibility curves. The absence of glassiness is explained
in more detail on Appendix~\ref{ExplainLogSG}, where we expose how the pair
correlation profile helps us in defining an appropriate susceptibility for the case
at hand, which is shown to diverge in the thermodynamic limit. It turns out that  this reflects not 
the existence of true glassiness but a transition closer to conventional ordering. 
Note that the gross features of the correlations (Fig.~\ref{PCvsXYfromMC} lower panel)
follow if one frustrates the pairs at the kink (Fig.~\ref{Int}) of the Log interaction,
which form a frame at half the system size. The set of points which in turn are on the ``frames''
of $O(L)$ points on the first frame yield the cross shaped set of ferromagnetically correlated
sites centred on the origin.

Note that such finite-size differences appear to be absent in previous studies in 
$d = 1$~\cite{larson2013spin}; they appear to be a consequence of the anisotropic
nature of our periodised Log interaction with its non-analytic minimum at maximum separation. 
By contrast, the ``smoothed Log''
(Fig.~\ref{Int}) that also respects the periodic boundary conditions essentially
reproduces the LGF interaction results.

\subsubsection{The fully covered square lattice}
For completeness, we have also analysed the situation for a fully occupied lattice.
In this case we observe that the LGF interaction leads to conventional (N\'eel)
antiferromagnetic order, while the Log leads to a ``striped'' phase. This can be
understood from a theorem in Ref.~\onlinecite{giuliani2007striped} which states
that the ground state of the system is determined by the minimum of the Fourier
transform of the interaction. This is explained in more detail on
Appendix~\ref{ExplainFullLatt}.

\subsection{Three dimensions}
\begin{figure}
\epsfig{file=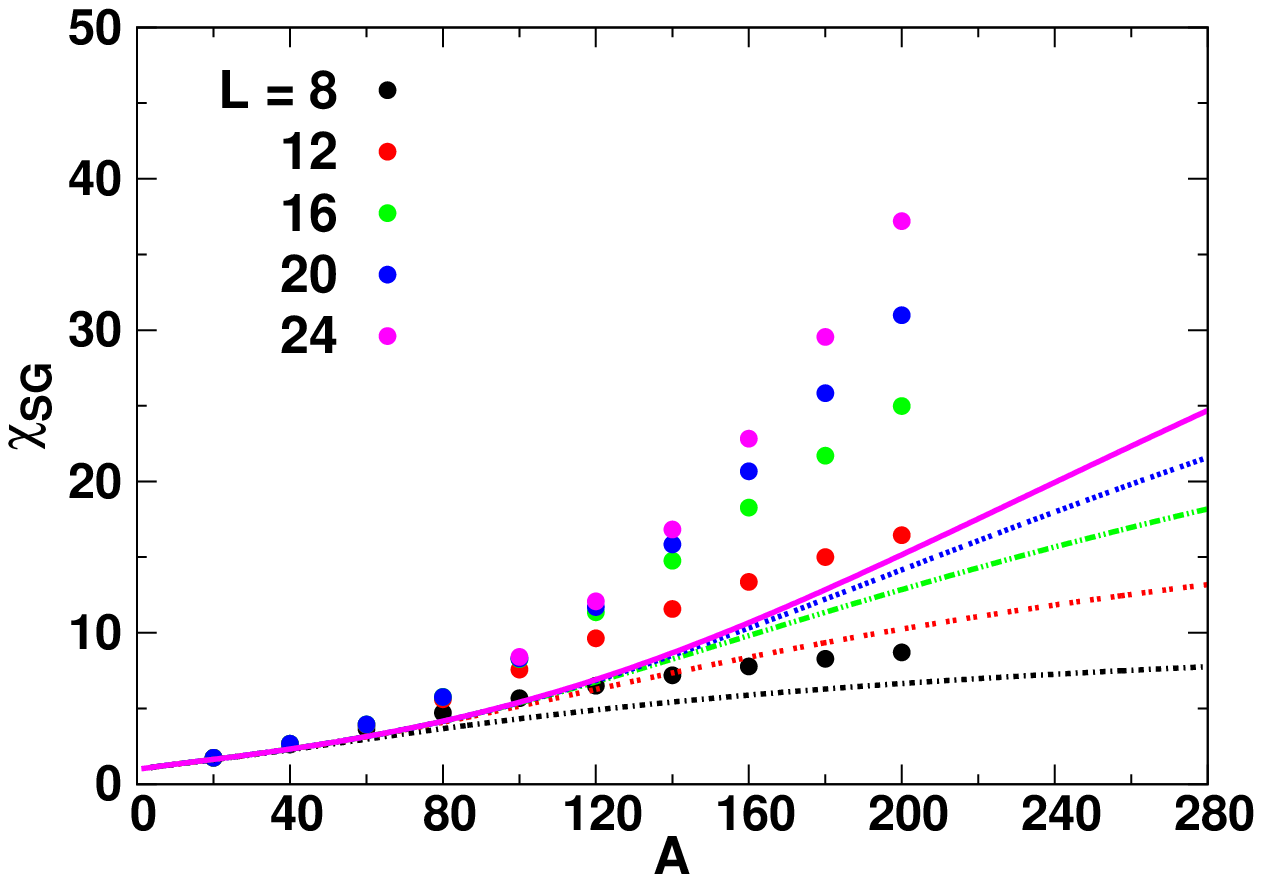,width=0.95\linewidth,angle=0} \\
\epsfig{file=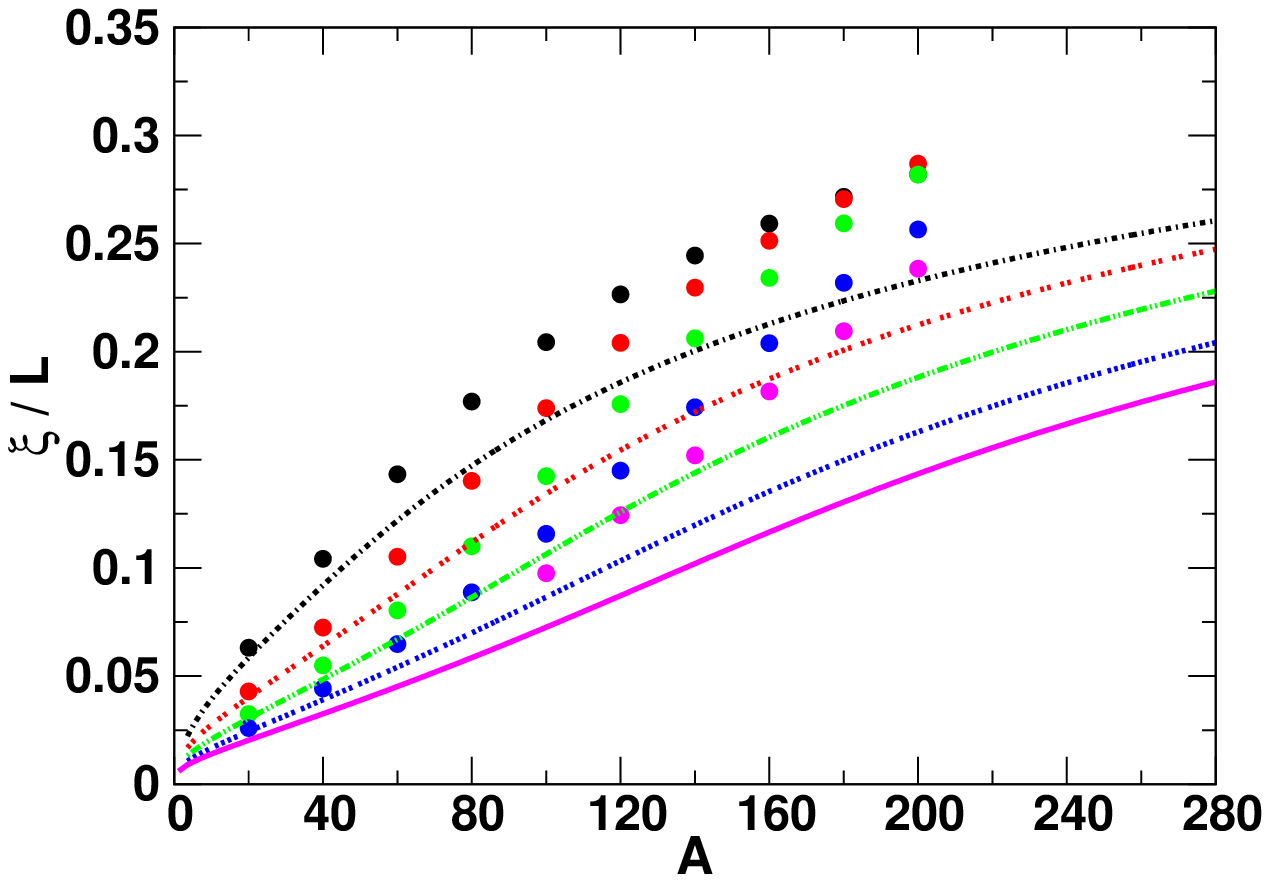,width=0.95\linewidth,angle=0}
\caption{Spin glass susceptibility (top) and correlation length (bottom) as computed from the
LM approach (lines) or measured in the MC simulations (points), for the LGF interaction
on the cubic lattice.}
\label{LMandMC3dLgf}
\end{figure}

We analyse the diluted cubic lattice considering a density of particles
$x=0.0625$, and again considering the model Hamiltonian of Eq.~\eqref{Hamilton},
with interactions now restricted to be the LGF as given by Eq.~\eqref{LGFDef}. Both Monte Carlo
simulations and LM calculations cover several system sizes with $100$ distinct disorder
realisations each. The main focus is on the possibility of a glassy phase
and the spin glass susceptibility and corresponding correlation length are computed.
Our prior discussion of the finite size scaling relations still holds, and one
determines the transition as an unique crossing of the finite size correlation length
curves. Instead of this we observe (Fig.~\ref{LMandMC3dLgf}) a trend for the crossings
to shift towards larger values of $A$ as the system size increases, similar to the
situation in two dimensions.

No good scaling collapse was obtained. A freezing transition in this system at a
finite coupling strength therefore appears unlikely, though a more careful finite size scaling
analysis of the crossings is necessary to give a definitive answer.

A LM study at $A=\infty$ reveals that the exponent for the  scaling of zero eigenvalues
of the matrix $B$ with system size yields $\mu=0.33$, in agreement with the prediction
in 3 dimensions for a short ranged interacting system~\cite{lee2005spin}. Using of this
exponent and the scaling relations at $A=\infty$ does not lead to a good scaling
collapse of our LM data, reinforcing the conclusion that this system does not present
any freezing transition at $A=\infty$.

The pair correlations exhibit the same sort of behavior as in the 2d case: only the
1st few nearest neighbors tend to be strongly antiferromagnetically correlated,
but no correlations develop at large distances as the coupling strength is increased,
and the system remains paramagnetic.
\begin{figure}
\epsfig{file=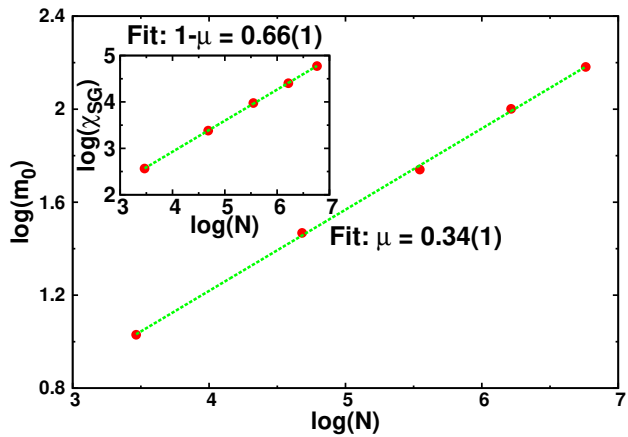,width=0.95\linewidth,angle=0} 
\label{LMandMC3dLgfScalM0}
\caption{Scaling of the number of vanishing eigenvalues of the matrix $B$
defined on the text and of the spin glass susceptibility (inset) with the number of
particles for the LGF in three dimensions at $1/A=0$.}
\end{figure}

\section{Spectral properties}
\label{RMT}

The $A^{-1}=0$ transition can be considered from the point of view of the interaction
matrix $J_{ij}$~\eqref{LGFDef} and~\eqref{LOGDef}, as an example of euclidean random
matrix (ERM):~\cite{mezard1999spectra} unlike the traditional random matrices, where
different entries of the matrix are uncorrelated, 
ERM's are defined by a function of the distance between two points $f(r)$, where the randomness 
in the entries is induced by the randomness of the underlying point pattern $\{\mathbf{r}_i\}$.
These random matrices have been studied for certain classes of functions $f$~\cite{goetschy2013euclidean},
and some classical results are available. Our degree of understanding of this subject is not
comparable to that of the classical (e.g.\ GOE,GUE, Wishart) ensembles~\cite{mehta2004random}
with most results coming from exact diagonalisation and approximations~\cite{goetschy2013euclidean,mezard1999spectra,amir2010localization}.

Unfortunately due to the long-range nature of the $\log$-interaction, many of the methods
to analyse the spectral properties presented in Ref.~\onlinecite{goetschy2013euclidean}
do not apply directly to our case. However, a phenomenological picture of the low-
and high-lying eigenstates of the matrix $J_{ij}$ can be established transparently.

\begin{figure}
\epsfig{file=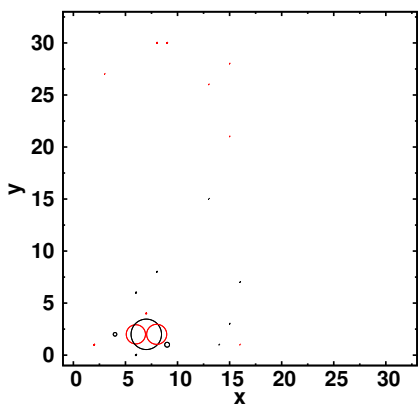,width=0.89\columnwidth,angle=0}
\caption{Ground state eigenvector showing a trimer for a particular disorder realisation
using the LGF as interaction on a lattice of size $L=32$ with $N=102$ particles.
The components of the eigenvector are proportional to the radius of the circles, which
are centered on the corresponding spin position. Red (black) sign indicates a positive
(negative) sign.}
\label{GSevEx}
\end{figure}

Let us start from the large positive eigenvalues. Since $J_{ij}$ is constant in sign,
the Frobenius-Perron theorem states that a highest eigenvector is nodeless. To a reasonable
approximation, it is fully delocalised,
\begin{equation}
\phi^{(N)}\simeq(1/\sqrt{N},...,1/\sqrt{N}).
\end{equation}
The associated eigenvalue is
\begin{equation}
\lambda_\text{max}\sim \frac{N}{2}\ln N.
\end{equation}
with an inverse participation ratio of $1/N$.

\begin{figure}
\epsfig{file=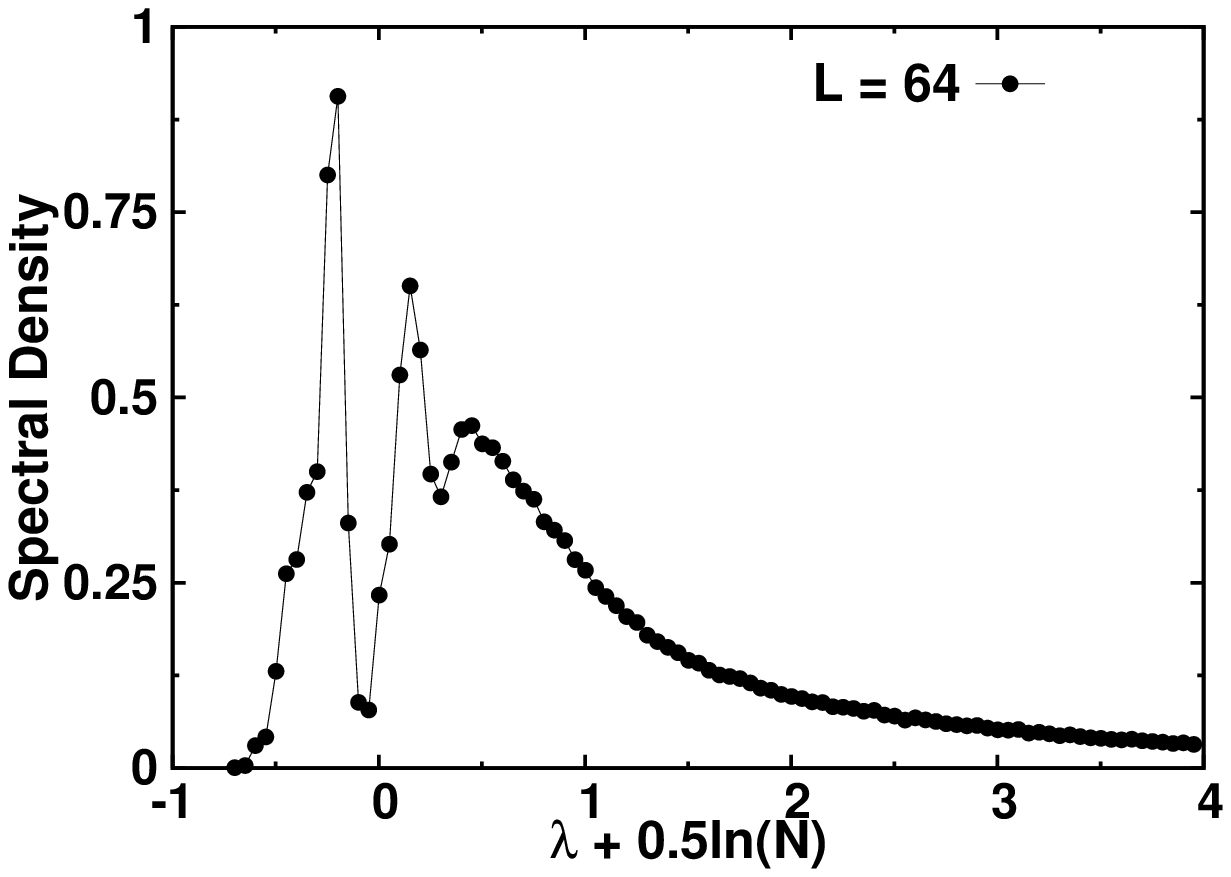,width=0.9\columnwidth,angle=0}
\epsfig{file=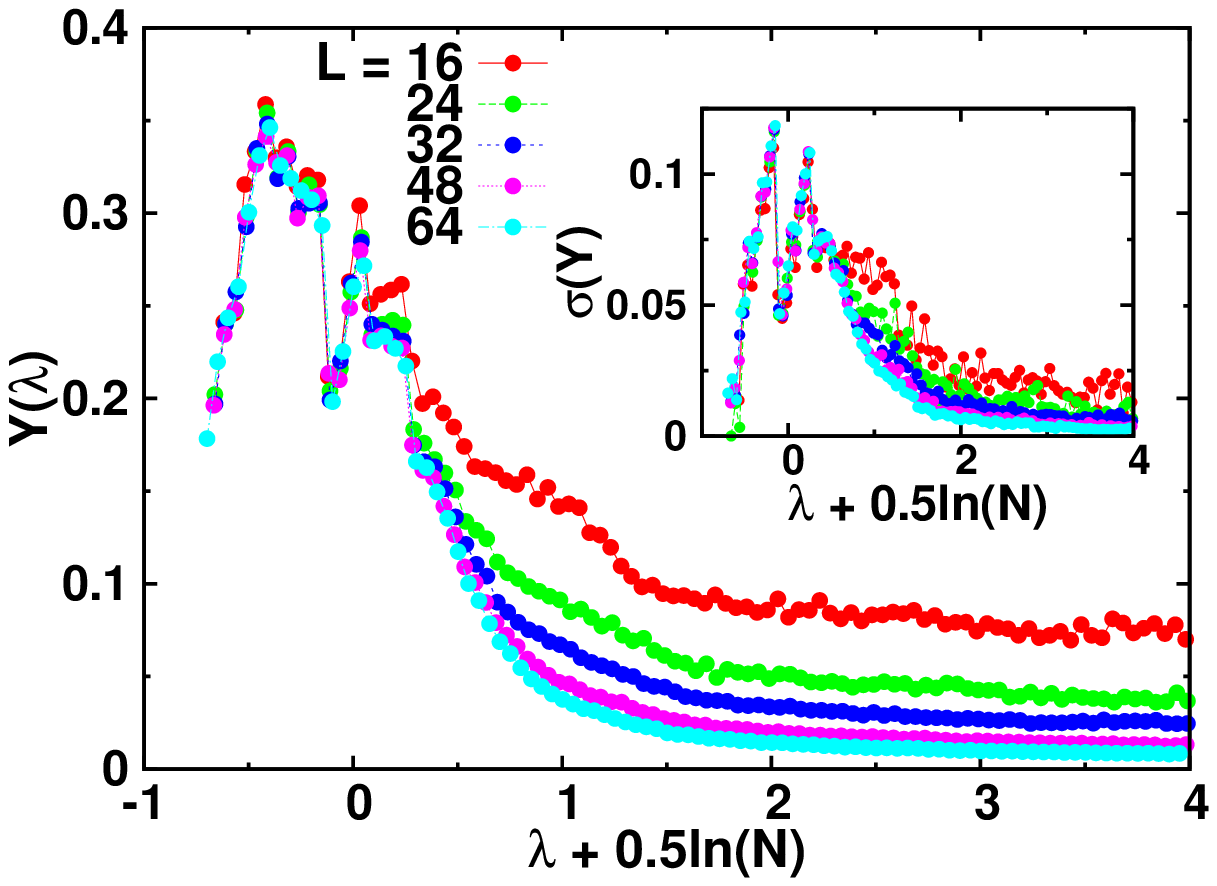,width=0.9\columnwidth,angle=0}
\caption{Spectral density (top) and average $Y$ (Eq.~\ref{eq:Y-lambda}, bottom) for a fraction
of $x = 0.125$ occupied sites in the lattice, using $J_{ij}$ as defined in~\eqref{LGFDef},
the LGF interaction. The inset shows the fluctuations of $Y$.}
\label{dosiprHighDens}
\end{figure}

\begin{figure}
\epsfig{file=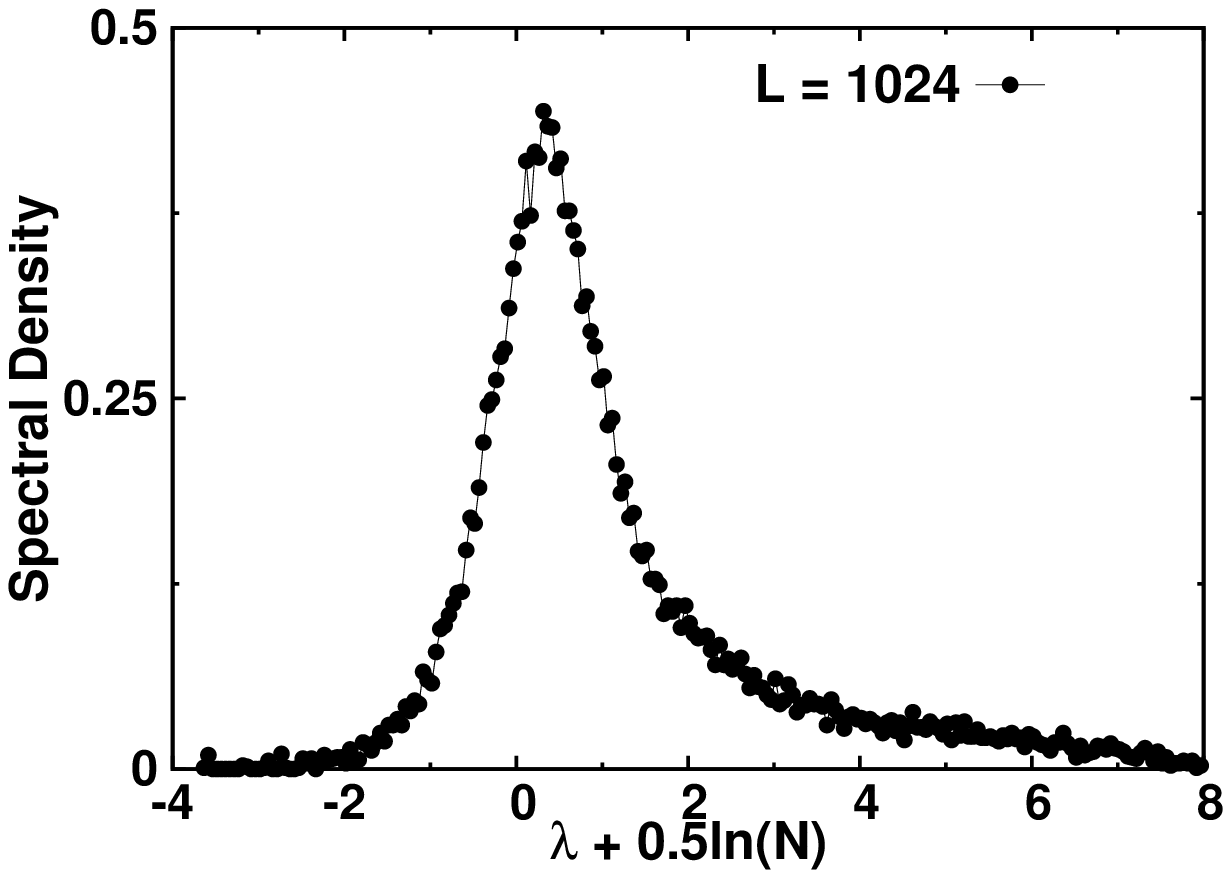,width=0.9\columnwidth,angle=0}
\epsfig{file=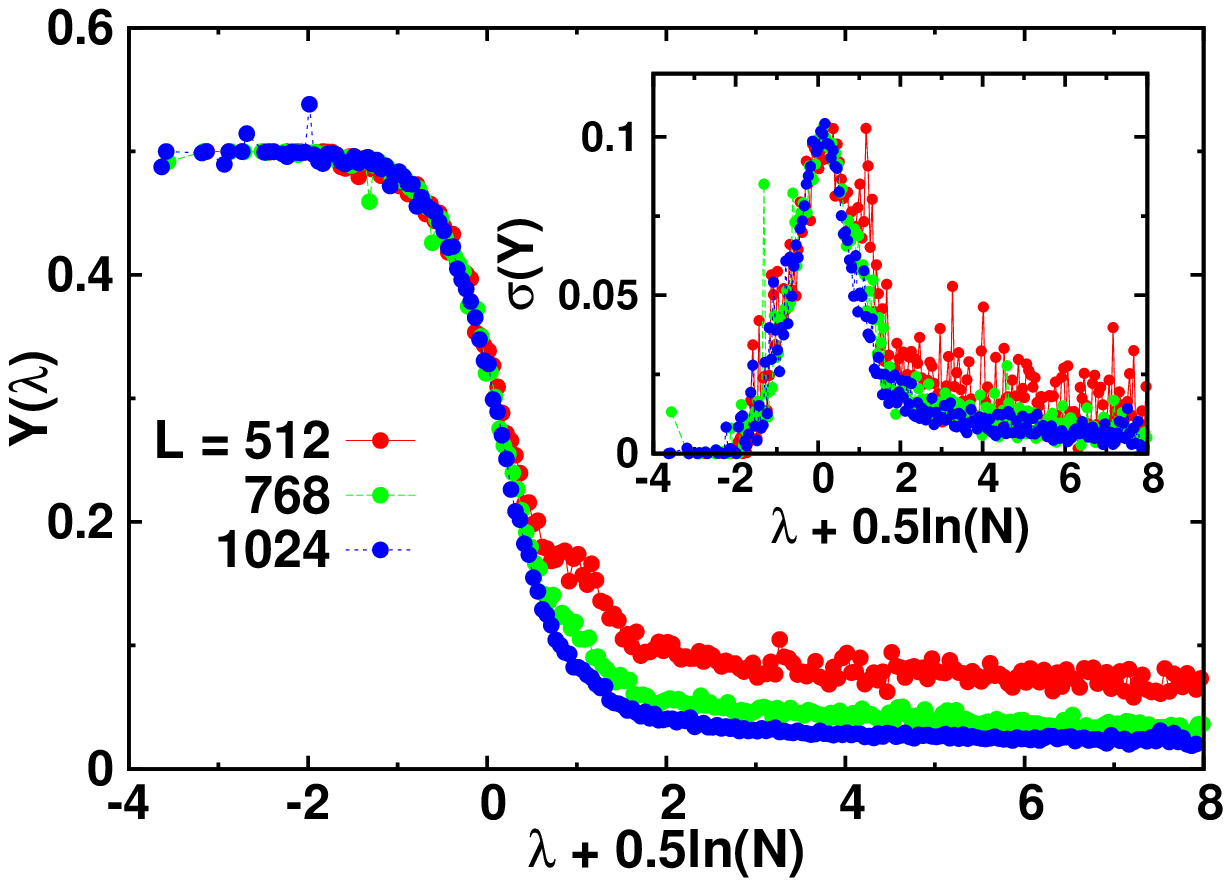,width=0.9\columnwidth,angle=0}
\caption{Spectral density (top) and average $Y$ (bottom) for a fraction of $x = 2^{-13}$ occupied sites
in the lattice, using $J_{ij}$ as defined in~\eqref{LGFDef}, the LGF interaction. The inset shows
the fluctuations of $Y$.}
\label{dosiprLowDens}
\end{figure}

The second-to-highest eigenvalue is also associated to a delocalised eigenvector, which is now
a wave with wavelength $\lesssim L$. At these length scales the randomness of the point
process plays little role.  A finite fraction (possibly all) of the eigenstates containing the largest
eigenvalues are \emph{delocalised}, they correspond to long-wavelength charge-density variations.
The average spectral density (DOS) of the LGF~\eqref{LGFDef} interaction matrices is shown
on top panels of Figs.~\ref{dosiprHighDens} and~\ref{dosiprLowDens}, in the limits of high
($x=0.125$) and low density (`continuum limit', $x=2^{-13}$), respectively.

Guided by the numerics, we see that the eigenvectors corresponding to the most
\emph{negative} eigenvalues are localised eigenvectors: most of the weight is concentrated
on $O(1)$ spins. This leads us to consider isolated percolation animals.

The simplest (and, for small $x$, the most abundant) of these is the dimer. A well-isolated dimer
supports two eigenvalues: an antisymmetric and a symmetric one. The antisymmetric one,
\begin{equation}
\phi^{(0)}=(1/\sqrt{2},-1/\sqrt{2},0,...,0)
\end{equation}
corresponds to the smallest eigenvalue. In fact, since the closest pair is located
one lattice spacing away $J_{12}\sim \ln L$ and the lowest eigenvalue is
\begin{equation}
\lambda_{\text{min}}\simeq-\ln (L)+O(1)\simeq \frac{1}{2}\ln(N/L^2)-\frac{1}{2}\ln(N)+O(1).
\end{equation}
At fixed density, $N/L^2$, the lowest eigenvalue depends logarithmically on the
system size.

For a well isolated dimer, say at distance $r$ from the closest spin, the effect of
neglecting the rest of the spins appears as a correction $O(1/r)$.

We now consider how big this isolation distance $r$ is. By the usual arguments
of percolation theory, one can estimate the expected number of isolated dimers as
\begin{equation}
n_d(r)=L^22x^2(1-x)^{\pi r^2},
\end{equation}
where we have approximated the number of lattice sites in a circle of size $r$ with $\pi r^2$.
Therefore the most isolated dimer (the solution of the equation $n_d(r)=1$) is surrounded
by an empty area of size
\begin{equation}
r(L)=\frac{\sqrt{2\ln(xL\sqrt{2})}}{\sqrt{\pi\ln(1/(1-x))}}.
\end{equation}
Note the extremely slow dependence $r(L)\sim\sqrt{\ln L}$.

Inserting $L=120$ and $x=0.1$, which is about the largest sizes considered in our numerics,
$r=4.13$, which can hardly be called isolated.

The isolation effect would be much more pronounced for $x=10^{-3}, L=1,200$, for which $r=18.4$.
Otherwise, one needs to consider the ground states of more complicated lattice animals,
like trimers, snakes, squares etc.
As an example, a ground state eigenvectors for one disorder realisation is shown on
Fig.~\ref{GSevEx}.

This problem becomes quickly analytically prohibitive. However the fact that the
ground state is localised on some lattice animal appears robust: on the graphs
we consider, the smallest eigenvalue is $\sim -\ln L$ and the IPR is $O(1)$.

With the lower end of the spectrum localised and the high-end delocalised, it is
a natural question whether there exists a mobility edge separating the two limits.
In order to study the transition we have looked at the inverse participation ratio as
a function of the eigenvalue $\lambda$:
\begin{gather}
	\text{IPR}_\alpha = \sum_i v_{\alpha i}^4\notag\\
	\label{eq:Y-lambda}
	Y(\lambda) = \frac{1}{\rho(\lambda)}\sum_\alpha \text{IPR}_\alpha \delta(\lambda - \lambda_\alpha),
\end{gather}
where $\lambda_\alpha$ and $v_{\alpha i}$ are eigenvalues and normalized eigenvectors of $J_{ij}$
respectively. We consider the average $[Y](\lambda)$ and fluctuations $\sigma(Y)(\lambda)$.~\cite{endnote1}
A mobility edge would be signaled by the divergence of the fluctuations of
$Y(\lambda)$ at a certain $\lambda_c$. Numerical diagonalization of $J_{ij}$ does
not indicate such a transition:
the two limits appear to be separated by a crossover. The bottom panels on Figs.~\ref{dosiprHighDens}
and~\ref{dosiprLowDens} show, respectively for a high and low density of particles,
the average $Y(\lambda)$, while the insets display the fluctuations of $Y(\lambda)$.
The spectral properties of the LGF in $d=3$ turn out to be very similar to the
$d=2$ case (not shown).

A detailed study of this ERM ensemble would be desirable and is left for future work.

\section{Pair correlations and screening}
\label{Screening}

\subsection{Analytical theory of screening}

Away from the $T\rightarrow0$ limit of the microscopic model,  excitations of the non-orphan 
tetrahedra  out of their momentless state carry gauge charge, which leads to a variant of Debye screening, 
with the special feature that the gaplessness of the charge excitations leads to a somewhat 
unusual temperature dependence of the screening length~\cite{sen2013coulomb}.

In addition to this, even in the limit $T\rightarrow0$ studied here, we encounter an additional type of 
screening. This occurs on account of the long-range uniformly antiferromagnetic Coulomb
interaction between the orphan spins, whose existence is the distinguishing property of
the random Coulomb antiferromagnet. It again exhibits a Debye
form, although distinct from the setting of mobile charges
in which Debye screening is normally considered, as here it is the (continuous) flavour
 of the charges -- the orientation of the orphan spin whose orientation is free but whose location is fixed -- which
is the dynamical degree of freedom. 

\begin{figure}
\epsfig{file=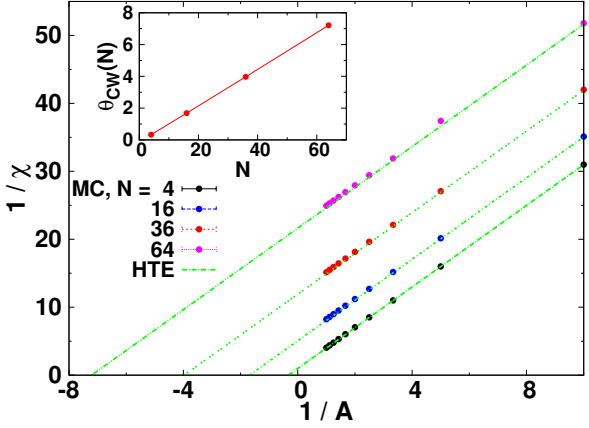,width=0.95\linewidth,angle=0} 
\caption{Uniform susceptibility as computed from MC simulations (dots) compared to a
weak coupling expansion averaged
over $200$ disorder realisations (dashed lines). The Curie-Weiss constant increases
approximately linearly with the number of particles (inset), yielding a vanishing radius of convergence
of the weak coupling expansion already at leading order.}
\label{MCandHTEUnifSusc}
\end{figure}

This can be seen directly in a weak-coupling expansion, which in Coulomb systems has a
vanishing radius of convergence in the thermodynamic limit, as is easily verified in our
simulations, Fig.~\ref{MCandHTEUnifSusc}.

To elucidate the role of screening, we compute the disorder averaged correlator
between two spins at $r_a$ and $r_b$. Consider the Hamiltonian
\begin{equation}
H=\frac{\alpha}{2}\sum_{i,j}J_{ij}\vec{n}_i\cdot\vec{n}_j,
\end{equation}
where $J_{ij}$ are given by either the Log or the LGF and we will eventually set
$\alpha=1$. The correlation function between two spins, for fixed disorder is:
\begin{eqnarray}
&C_{ab}=\la\vec{n}_a\cdot\vec{n}_b\ra=\nonumber\\
&=\frac{1}{Z}\int d^{3N}n\prod_i\delta(1-n_i^2)(\vec{n}_a\cdot\vec{n}_b)e^{-\frac{\alpha}{2}\sum_{i,j}J_{ij}\vec{n}_i\cdot\vec{n}_j}.
\end{eqnarray}

As it is not the hard spin constraint which is central to the physics of screening, we substitute it
with something more manageable (analogously to the LM method, but without imposing
self-consistency). 
Representing the delta function with a Gaussian term
\begin{equation}
\delta(1-n_i^2)\to\frac{1}{(2\pi/3)^{3/2}}e^{-3\frac{n_i^2}{2}}
\end{equation}
(with a factor of $3$ to guarantee that $\la n^{x2}_i+n^{y2}_i+n^{z2}_i\ra=3/3=1$). Thus
\begin{equation}
\label{eq:Crab}
C_{ab}=\delta_{ab}-\la a|\frac{\frac{1}{3}\alpha J}{1+\frac{1}{3}\alpha J}|b\ra
\end{equation}
where we use a matrix notation $\la a|J|b\ra=J_{ab}$. For simplicity we will not
write the $\delta_{ab}$ term, which only affects the result for the self-correlation
(it will return to be important when we discuss the LM approximation again later). The
correlation function between $a$ and $b$ depends also on the positions of all the other
points $x_2,...,x_N$ so it should be written as $C(x_a,x_b|x_2,...,x_N)$.

This Gaussian approximation is equivalent to the resummation of a set of diagrams in
which there are no internal loops, dubbed ``chain diagrams." This approximation is
justified in the limit of small $\alpha$, in which spins are rarely polarized along some
direction and the hard-spin constraint is not so important.

This result holds for each disorder realization. We now take the average over
realizations (leaving the question of whether this is representative of the
distribution or not for later) keeping fixed the position of the two spins
$a,b$. For doing this, it is convenient to go back to the geometric expansions
and define
\begin{equation}
\avg{C_{ab}}\equiv\int \frac{d^{N-2}x}{S^{N-2}}C(x_a,x_b|x_1,...,x_{N-2})
\end{equation}
where $x_i$ are the locations of the other $N-2$ spins and $S=L^2$. We have relaxed the
constraint that points be located on a square lattice, which is immaterial in our high
temperature, low-dilution expansion.

Unfortunately it is difficult to see what the distribution of $J$ induced by the random
positions is, but we can expand the Gaussian result in powers of $\alpha$ and do the
average term by term.

We get
\begin{eqnarray}
\avg{C_{ab}}&=&-\frac{1}{3}\alpha J_{ab}+\sum_{i}\left(\frac{1}{3}\alpha\right)^2\avg{J_{ai}J_{ib}}\nonumber\\
&-&\left(\frac{1}{3}\alpha\right)^3 \sum_{ij}\avg{J_{ai}J_{ij}J_{jb}}+...
\end{eqnarray}
Now, term by term we obtain objects like
\begin{eqnarray}
\avg{\sum_i J_{ai}J_{ib}}&=&(N-2)\int\frac{d^2x}{S}J(x_a-x)J(x-x_b)\nonumber\\
&=&\rho\int d^2x J(x_a-x)J(x-x_b)
\end{eqnarray}
where $\rho=(N-2)/S\simeq N/S$ is the density of points. Fourier transforming,
\begin{eqnarray}
\rho&\int &d^2x J(x_a-x)J(x-x_b)\nonumber\\
&=&\rho\int d^2x \frac{d^2q}{(2\pi)^2} \frac{d^2q'}{(2\pi)^2} J_q J_{q'} e^{iq(x_a-x)+iq'(x-x_b)}\\
&=&\rho \int\frac{d^2q}{(2\pi)^2}J_q^2 e^{iq(x_a-x_b)}.
\end{eqnarray}
The geometric series obtained thus for $\avg{C_{ab}}$ yields
\begin{equation}
\avg{C_{ab}}=-\int \frac{d^2q}{(2\pi)^2}e^{iq(x_a-x_b)}\frac{(\alpha/3)J_q}{1+(\alpha\rho/3)J_q}.
\end{equation}
Now, for both Log and the LGF, 
$J_q\simeq c/q^2$  ($c$ is a constant of $O(1)$)~\cite{endnote2} 
so that at small $\alpha$ we have approximately
\begin{equation}
\avg{C_{ab}}\simeq-\int \frac{d^2q}{(2\pi)^2}e^{iq(x_a-x_b)}\frac{(c\alpha/3)}{q^2+(c\alpha\rho/3)}\ .
\end{equation}
This leads to 
\begin{equation}
\avg{C_{ab}}\simeq(-2\alpha/3) K_0(r\sqrt{c \alpha \rho/3})
\end{equation}
which exhibits a screening length
\begin{equation}
\label{eq:corlen}
\xi=1/\sqrt{c\alpha\rho/3}.
\end{equation}
As both $\alpha$ and $c$ are $O(1)$ this shows (not surprisingly) that the screening length
is proportional to the $1/\sqrt{\rho}$. 

\begin{figure}[htbp]
\begin{center}
\epsfig{file=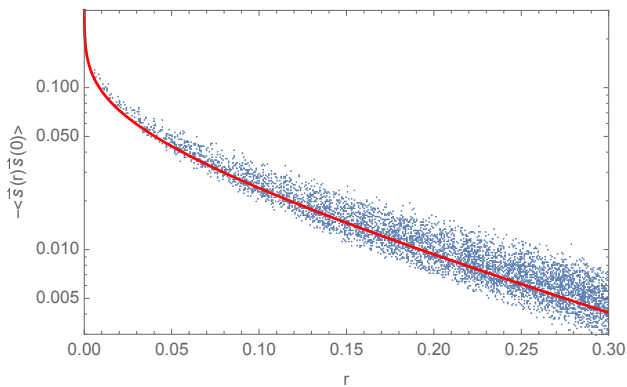,width=0.95\linewidth,angle=0} 
\caption{Correlation function exhibiting screening: numerical results (for a
single disorder realization with $N=200$ points on a square of unit size)
compared  to the predicted analytical form from the chain diagrams.}
\label{default}
\end{center}
\end{figure}
Note that in this approximation, for $r_{a,b}\ll \xi$ the
correlation function $C(r)\gg 1$, which is not physical for unit length spins.
This is an artefact resulting from substituting the  hard spin constraint with a quadratic
confining potential. Therefore this approximation is internally consistent only for $r_{a,b}\gtrsim \xi$,
where it predicts an exponential damping of the correlations but we note that the large {\it anti}correlations
at short distance due to strongly coupled spins close to one another put these into a state with vanishing
total spin, which -- physically correctly -- screens their joint field at larger distances. 

\subsection{A random scattering picture}

The final question we address concerns the fluctuations of the random quantity~\eqref{eq:Crab} and
whether these may signal any  phase transition even when the mean does not. 
To gain some insight into this, we develop
an analogy with wave propagation in
disordered media, which suggests that no transition exists.
The basic observation is that the interaction is simply related to the inverse of the Laplacian,
the propagator of a free particle on the lattice:

Considering that
\begin{equation}
J_{ij}=\la i|\frac{1}{-\nabla^2}|j\ra
\end{equation}
properly regularized (particularly important is the condition that $J_{ii}=0$), we can rewrite
the expression~\eqref{eq:Crab} as 
\begin{equation}
C_{ab}=\delta_{ab}-\frac{\alpha}{3}\la a|\frac{1}{-\nabla^2+V-E}|b\ra,
\end{equation}
where $E=0$ and
\begin{equation}
V(x)=\frac{\alpha}{3}\sum_{i}\delta(x-x_i)
\end{equation}
is a random potential. This can be established by expanding  in powers of $\alpha$.

Thus $C$ is (proportional to) the propagator for a wave in a two-dimensional box with randomly
placed point-like scatterers~\cite{albeverio2005solvable,ishimaru1991wave}, at energy $E=0$. 

The precise form of the mapping is the following: the correlation function
\begin{equation}
-\frac3\alpha\la n_an_b\ra,
\end{equation}
is the amplitude of a signal sent from the scatterer $a$ to the
scatterer $b$, considering all order processes bouncing over all the $N$ scatterers. In case
$a=b$ the direct path from $a$ to $b$ needs to be neglected. This is a form of
\emph{renormalization} of the scattering problem which is always necessary in the
point-like (or $s$-wave) scattering limit~\cite{scardicchio2005casimir}.

Once the renormalization procedure is done, the problem we are left with corresponds
to the propagation of a scalar wave, damped by a scattering section for every typical
realization of disorder. Without repeating the classical treatment of this phenomenon
we can say that the signals (spin-spin correlations) must be screened for any $\alpha$,
the screening length (measured in units of $1/\sqrt\rho$)
being a decreasing function of $\alpha$. Even if not precisely of the form~\eqref{eq:corlen}
for small-$\alpha$, it seems to diverge like $1/\sqrt\alpha$.

This is valid both for the \emph{coherent} field $\avg{C_{ab}}$ and the \emph{incoherent}
field $\avg{C_{ab}^2}-\avg{C_{ab}}^2$, although the scattering sections (and hence
the damping/correlation lengths) might have different values. This analogy makes us
realize that in this approximation there is \emph{no transition} irrespective of the
value of $\alpha$ or $\rho$, and this is consistent with numerical results.

This analogy extends also to the LM limit. Considering a small-$\alpha$ series
expansion for the spin correlation function:
\begin{equation}
h_aC_{ab} h_b=\delta_{ab}h_a-\alpha J_{ab}+\alpha^2J_{ai}\frac{1}{h_i}J_{ib}-\alpha^3J_{ai}\frac{1}{h_i}J_{ij}\frac{1}{h_j}J_{jb}+...
\end{equation}
(recall that in LM $\alpha$ is scaled by a factor $1/m$,
hence the factor of 3 of the previous paragraphs is absent here) where
the extra factors of $h_a$ need to be chosen
in such a way that
\begin{equation}
C_{ii}=\la n_in_i\ra=1.
\end{equation}
$C_{ab}$ is then proportional to the propagator 
\begin{equation}
G_{ab}=\tilde{h}_{ab}-\alpha\la a|\frac{1}{-\nabla^2+V-E}|b\ra,
\end{equation}
where $\tilde h$ is the diagonal matrix with diagonal entries $\{h_i\}_{i=1,...,N}$, $E=0$ and 
\begin{equation}
V(x)=\alpha\sum_{i}\frac{1}{h_i}\delta(x-x_i),
\end{equation}
where the renormalized value $\la i|\frac{1}{-\nabla^2}|i\ra=0$ is intended.

This is a scattering problem over point-like scatterers, where now each scatterer
has different scattering amplitude. This modification should not change the physical
analogy of the problem. This is again a scattering problem of a scalar wave over
point-like scatterers. The propagation of the wave is attenuated over distance in
the usual exponential fashion. Therefore, if a phase transition exists, it is not
mirrored in the divergence of the correlation length. Conversely, as this treatment
is closely related to the LM one (rather than the Heisenberg model), on account of
the softening of the hard constraint to a Gaussian one, we would not expect a 
transition at finite value of $\alpha$.

\section{Discussion}

We have studied the effective theory describing 
disorder in the form of quenched non-magnetic impurities, in the topological Coulomb
phase, on a lattice with bipartite dual. Interactions in the effective picture are
long-ranged, and to the best of our knowledge this is the first study available of
such a model.

\subsection{A freezing transition?}

Our results show that any freezing transition, if it exists, is
extremely tenuous. In $d=2$, for LM there does not appear to be
freezing for any finite coupling, with a nice scaling collapse of the
data at $A=\infty$ indicating freezing to take place in this limit.

The relation of this result to a finite number of spin components
is the following. Firstly, our Heisenberg simulations cannot access a
freezing transition, but they do show a greater tendency towards
glassiness than LM, with both a larger spin-glass correlation length
and an enhanced tendency for the curves to cross. 

This is in keeping with the general expectation~\cite{lee2005spin} for
the more constrained Heisenberg model to freeze before the soft spins
do (anf after an Ising model might). If there is a freezing transition
at $A_c<\infty$, it will still be at phantastically large coupling
$A_c>100$.  The delicate nature of all of these phenomena is further
underscored by the dependence on finite-size choices, which may lead
to an entirely different set of instabilities. Similarly, the analytical approaches, 
in particular the mapping to a quantum scattering problem, see little
indication of a transition. 

The tendency towards freezing seems to be even weaker in $d=3$,
perhaps surprisingly so, given the freezing transition is more robust
in higher dimension for the instances of canonical spin
glasses. However, unlike in these cases, our distribution of the
intersite couplings is dimensionality dependent, and in particular
becomes 'shorter-ranged' as the power law of the decay of the Coulomb
law grows with $d$ (while, of course, the power law with which
the number of distant spins grows, increases). 

The weak tendency towards freezing is in keeping with the fact that
our model is not easily deformed into one of the standard spin glass
models. On one hand, increasing the range of the interaction towards
the extreme of doing away with any notion of distance and assigning
equal coupling between all the spins yields simply a global
charge-neutrality constraint (which, at any rate, is already enforced
microscopically) and therefore preserves a microcanonical version of a
{\em perfect paramagnet}. If the coupling is restricted to 
nearest-neighbours only, we instead get a combination of percolation
physics and that of the  standard N\'eel state for 
a bipartite antiferromagnets, where  any tendency towards
disorder is a dimensionality effect, and glassiness is nowhere to be
seen.

The tendency towards glassiness is therefore necessarily due to a
combination of the non-constancy of the logarithmic interaction --
which, helpfully, is not bounded as $r\rightarrow\infty$, along with
its long range. Studying models exhibiting this pair of ingredients
more systematically is surely an interesting avenue for future
research. We would like to emphasize, in particular, that the
phenomenon of screening we have discussed has no counterpart in the
literature on conventional spin glasses, where the random choice of
the sign of the interactions does not allow the identification of an
underlying charge structure.

In this sense, our model is much closer to those familiar from the
study of Coulomb glasses, although the differences here are again
considerable. We have vector charges rather than Ising (positive or
negative) ones; disorder appears in the form of random but fixed {\it
locations} rather than fixed on-site potentials for charges not bound
to a particular site. It is intriguing that such a variation of a
classic Coulomb glass appears entirely naturally in frustrated
magnetism. 

\subsection{Freezing in frustrated magnetic materials}
\label{DiscFreezInFrust}

With Heisenberg spins placed at random sites of the pyrochlore-slab lattice
(also known as the SCGO lattice) and a particular, microscopically determined
value of $A$, the $d=2$ case of our Coulomb antiferromagnet corresponds, up
to the sublattice-dependent  inversion factor mentioned earlier, to the
$T\rightarrow 0$ limit of the physics of orphan-spins created when a pair
of Ga impurities substitutes for two of the three Cr spins in a triangular
simplex of this lattice. Although experimental interest in SCGO dates back to
the 80s and played a key role in stimulating experimental and theoretical interest in the area
of highly frustrated magnetism\cite{obradors1988magnetic}, the behaviour 
of SCGO is reasonably well-understood in theoretical terms only
in the broad Coulomb spin-liquid regime down to about a hundredth of the
exchange energy scale (of order 500K) between the Cr spins. The magnetic response in this regime
can be modeled in a rather detailed way as being made up as the response
of a pure Coulomb spin-liquid superposed with the Curie-tails associated
with vacancy-induced ``orphan-spin'' degrees of
freedom~\cite{schiffer1997two,moessner1999magnetic,Henley2001,Henley2010coulomb}
that carry an effective fractional spin~\cite{SenDamleMoessnerPRL,sen2012vacancy}
and leave their imprint on NMR lineshapes~\cite{SenDamleMoessnerPRL} and
bulk-susceptibility~\cite{schiffer1997two,moessner1999magnetic,SenDamleMoessnerPRL}
in the Coulomb spin-liquid phase. In contrast, the physics at very low temperatures
(of order 5K or lower) is still not very well understood, with intriguing but
largely unexplained reports of observed glassy behaviour even at very low
densities of Ga impurities\cite{ramirez1990strong, laforge2013quasispin},
which appears to involve only  the freezing of a  fraction of its degrees of freedom.

Our model retains the key feature of the $T\rightarrow 0$ limit of the
effective model, namely the long-range Coulomb form of the effective exchange
couplings, but does not retain the detailed geometry of these orphan-spins
in SCGO, except for the sublattice-dependent inversion that connects the
degrees of freedom of our Coulomb antiferromagnet with the underlying
physics of these orphan-spins.

Bearing all this in mind, the usual caveat about idealised models
for frustrated systems applies to our study as well: Our starting
Hamiltonian of a classical nearest-neighbour Heisenberg model does
not include a number of aspects -- further-neighbour interactions,
single-ion anisotropies, non-commutation of spin components -- all
of which give rise to interesting, generally non-glassy, physics of
their own. If and when these energy scales dominate over our the
instabilities of the idealised model, it is the former which will
likely show up more prominently in experiment. 

In addition, in our case, the critical coupling $A_c$, even if it is
not infinite, is hard to attain in any microscopic model. Indeed,
for the checkerboard lattice, one obtains {\bf $A=1/4\pi$} from a
microscopic calculation, easily within a very short-range correlated
regime.

At any finite temperature, which is all that can be accessed
experimentally for the time being, the Coulomb interactions obtain a
finite-screening length due to the thermal excitation of charges even
in non-orphan tetrahedra. Following the general lore on spin freezing,
this precludes even canonical Heisenberg spin glassiness. For this
reason, the abovementioned $A$-independence of a freezing transition
in $d=2$ is not going to carry over directly to the experimental
compound. 

However, real systems will only be quasi-$2d$, with residual
couplings between the two-dimensional layers. Indeed, for the case of
SCGO, dilution also breaks up the tightly bound singlets of the dimers
of Cr ions which isolate the kagome-triangle-kagome trilayers from one
another. The consequences of coupling in the third dimension remain an
interesting yet completely open topic for future study.

\begin{figure}
\hspace{-0.3cm}\epsfig{file=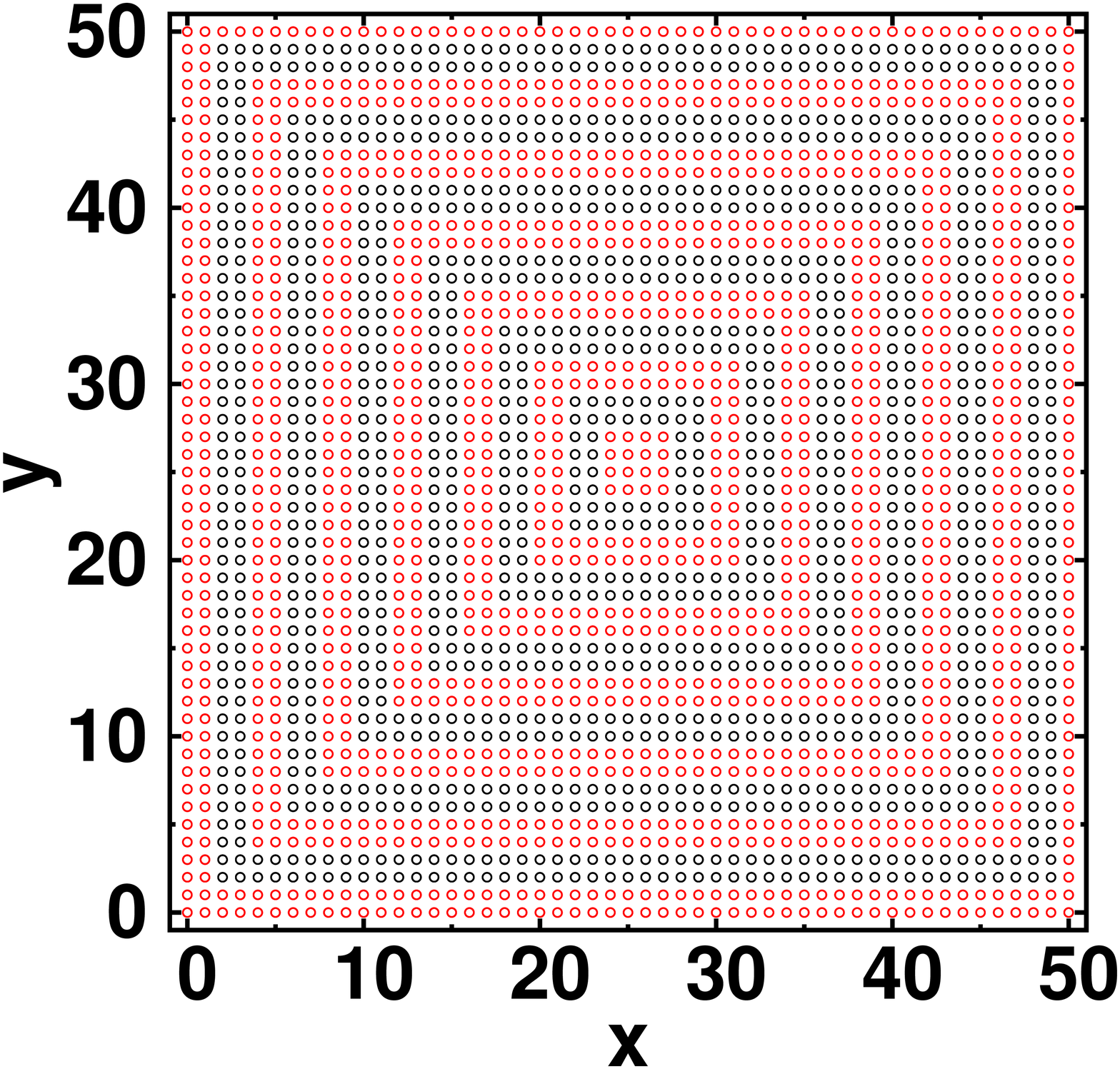,width=0.5\linewidth,angle=0}
\epsfig{file=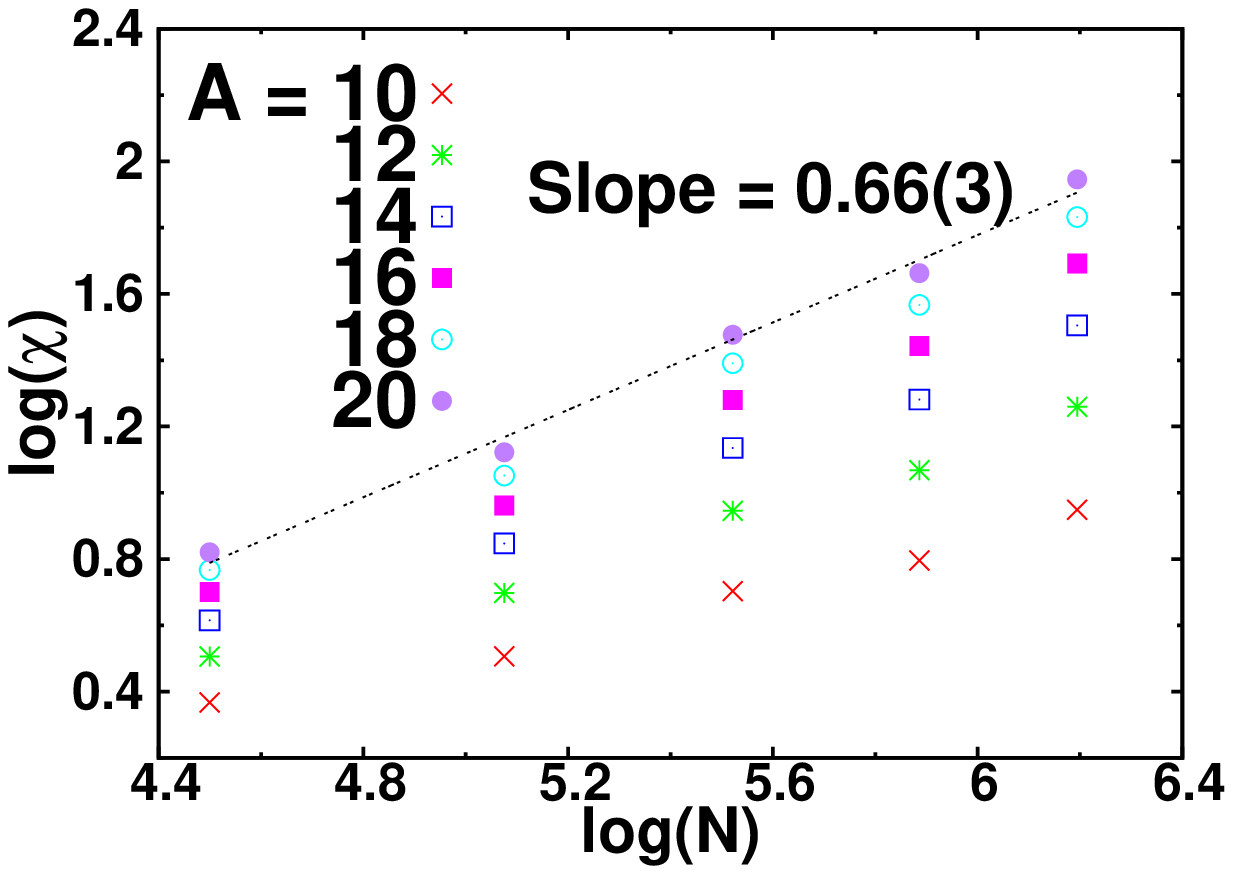,width=0.5\linewidth,angle=0}
\caption{(Left)The sign function on the 1st quadrant for a lattice of side $L=100$ used
to resum the correlations. (Right) The scaling of the new susceptibility proposed to
describe the ordering occuring with the Log interaction.}
\label{RedefChi}
\end{figure}

\begin{figure}
	\hspace{-0.3cm}\epsfig{file=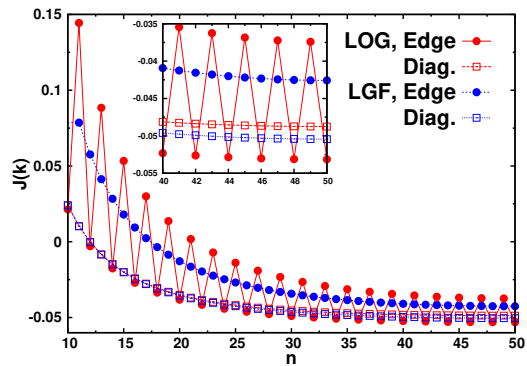,width=0.8\linewidth,angle=0}
\caption{Numerically obtained Fourier transform of Log (red) and LGF (blue) for
a lattice of size $L=100$. Circles connected by lines indicate the edge $k_y=0$, while
squares indicate the diagonal $k_x=k_y$. The index $n$ labeling the $x$ axis indicates
the index of the wave vector: $k_x=\frac{2\pi}{L}n$. The inset shows in more detail the
region near the global minimum.}
\label{FTInter}
\end{figure}
\begin{figure}
\hspace{-0.3cm}\epsfig{file=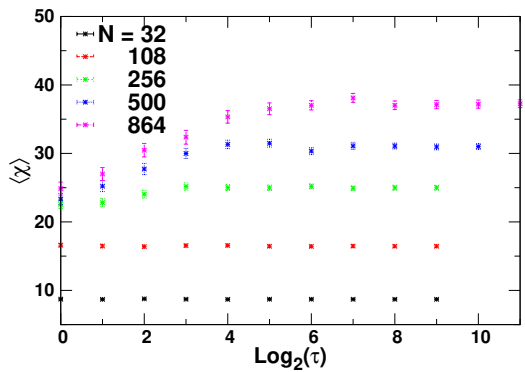,width=0.8\linewidth,angle=0}
\caption{The average on each binning block of the spin glass susceptibilities
plotted against the logarithm (base 2) of the size of the corresponding binning block.
Data shown here corresponds to MC simulation of the LGF in the cubic lattice at $A=200$.}
\label{FigTstEquil}
\end{figure}

\subsection{Connection to other models}
More broadly, perhaps the most pleasing aspect of this work is how it naturally connects (with)
a  number of deformations of well-known problems--the scattering problem, Coulomb glass physics,
or random matrix theory. In particular, we have identified a straightforward
way of obtaining a Euclidean random matrix problem from a simple magnetic model where long-range
interactions emerge naturally. We hope that this will motivate further work on any (and perhaps all)
of these problems. 

\section*{Acknowledgements:} We are very grateful to John Chalker, Ferdinand Evers,
Mike Moore and Peter Young for useful discussions. This work was supported by DFG
SFB 1143.

\appendix
\section{ Non-Glassiness for the Log Interaction}
\label{ExplainLogSG}
The pair correlation profiles for the Log interaction exhibit a structure hinting
on the way pair correlations should be summed in order to define a generalized
susceptibility describing the order present on this system. This order reflects the
symmetry of the interaction, which is anisotropic, but has the symmetries of the square
lattice.

We define a {\it sign function}, $\theta(x,y)$, which on each quadrant has
alternating values $\pm 1$ on suscessive ``square frames'' of fixed width of $2$
lattice sites for any $L$. Assuming $(x,y)$ on the
1st quadrant, this function has the profile pictured on the left panel of
Fig.~\ref{RedefChi}.

The corresponding susceptibility reads:
\begin{align}
\chi = \left[\frac{1}{N}\sum_{i,j} \theta(\vec{r}_{ij})\left<\vec{S}_i\cdot\vec{S}_j\right>\right] .
\label{NewChi}
\end{align}

Square brackets denote as usual disorder average.
This susceptibility diverges with system size, and its scaling in MC simulations
is shown on the right panel of Fig.~\ref{RedefChi}; the same behavior is
found in the LM.

\section{ Fully Occupied Lattice }
\label{ExplainFullLatt}

Proposition $1$ in Ref.~\onlinecite{giuliani2007striped} states that if $\hat{J}(k)$
is the Fourier transform of the interaction matrix $J$, then a minimizer $\vec{k}_0$
for $\hat{J}(k)$ determines a modulated ground state for the system with that wavevector.

The Fourier transform of the LGF at nonzero wavevector is readily read from its
definition, Eq.~\eqref{LGFDef}:
\begin{align}
	\hat{J}_{\text{LGF}}(k) = \frac{1}{2-\cos(k_x)-\cos(k_y)}
\end{align}
which has a minimum at $\vec{k}=(\pi,\pi)$, thence we find ``conventional''
antiferromagnetic order.

For the Log interaction, we are not able to find an analytical expression for its
Fourier transform, but numerical results show that the global minima happen at
$\vec{k}=(\pi, 0)$ or $(0, \pi)$, which explains the striped phase for the fully
occupied lattice. The non-analyticity of the distance function periodized by
the functions $\text{min}(x,L-x)$ or $\text{min}(y,L-y)$ (which is seen as a
discontinuity in the derivative along the lines $x=L/2$ or $y=L/2$) gives rise
to ``ringing'' in $\hat{J}_{\text{Log}}(k)$, a line of alternating local maxima
and minima appear along $k_x=0$ or $k_y=0$. The new global minimum is shifted
from $(\pi,\pi)$ to the edges of these lines, as shown in Fig.~\ref{FTInter}.

\section{ Verifying Equilibration}
\label{TstEquil}

Our simulations require exploring a region of very high coupling, $A$.
In this case it is important to ensure that equilibrium is attained.
To test this, we bin the data for the spin glass susceptibility. This binning
consists of subdividing the total number of measurements,
$N_{m}$, in contiguous bins of successive sizes: $1,1,2,4,8,\ldots,N_m/4,N_m/2$.
The average for each bin is then plotted against the logarithm of the bin size
(Fig.~\ref{FigTstEquil}). Equilibrium is diagnosed by at least the last $3$ bin
averages agreeing within the interval set by their error bars.

The final equilibrium values used consist of the average of the last half of the
measurements made in the simulation, $N_m/2$.

\bibliography{ManuscriptBib}

\begin{thebibliography}{34}
\bibitem[{end({\natexlab{a}})}]{endnote1}
\bibinfo{note}{Numerical evaluation of $Y$ reduces to the binning of the
  eigenvalues interval and computing the histogram of PRs falling into the
  bins. We defined the fluctuations as the sample-to-sample variance of the bin
  values.}
\bibitem[{end({\natexlab{b}})}]{endnote2}
\bibinfo{note}{In order to find the appropriate $c$ we need to specify $J$.
  From $$ c\int\frac{d^2q}{(2\pi)^2}\frac{e^{iqr}}{q^2+2/L}=
  \frac{c}{2\pi}\ln\left(\frac{L/\sqrt{2}}{r}\right)+\mbox{const.} $$ it
  follows that $J(r)=-\ln (r/L/\sqrt{2})$ implies $c=2\pi$.}
\end{thebibliography}
\end{document}